\documentclass[prd,preprint,12pt]{revtex4}
\pdfoutput=1
\usepackage{hyperref}
\usepackage{graphicx}
\usepackage{amsmath,amssymb,amsfonts,amsthm,latexsym,stmaryrd}
\usepackage{marginnote}
\usepackage{color}
\usepackage{soul}

\newcommand{\FK}{K}
\newcommand{\cK}{\lambda}
\newcommand{\cC}{\sigma}
\newcommand{\C}{A}
\newcommand{\D}{B}
\newcommand{\cR}{\kappa}  
\newcommand{\Lm}{L_m} 

\begin{document}   
\title{Tidal love number of neutron stars with conformal coupling}
\author{Hamza Boumaza}\email{boumaza14@yahoo.com, boumaza.hamza@univ-jijel.dz}    
\affiliation{\small Laboratory of Theoretical physics (LPTH) and Department of Physics, faculty of Exact and Comuter Science, Mohamed Seddik Ben Yayia university, Bp 98 Ouled Aissa, Jijel 18000, Algeria}
 \begin{abstract}
 In this present work, we suggest studying neutron star with a conformal coupling by using the analytical expressions  of realistic  equations of state. The equations of perturbed and nonperturbed metrics, scalar field and the pressure of the matter are derived in order to observe the effects of the conformal coupling on the mass, radius and tidal love number of polar and axial types. The numerical and the analytical analysis show a different comportment of our model from GR by varying the value of the parameters in the model. We showed also for a particular case that the scalar field becomes vanished outside the star which will give us the opportunity to investigate the behaviour tidal love number of our model.   Finally, a comparison between the results of our model and the  outcome from the signal of gravitational waves, coming from the fusion of binary neutron star, was performed using the universal relations $C$-$\Lambda$.
 \end{abstract} 

\maketitle

\section{Introduction}

{\hskip 2em} In the last century neutron stars (NSs) were viewed as the most astrophysical objects to probe the new  theories of gravity and  new equation of states (EoSs) through the gravitational waves (GWs) \cite{Abbott_2017,LIGOScientific:2017vwq}. In fact,  the detection of GWs from a collision binary neutron stars  or binary black holes-neutron star (BH-NS) marks the beginning new eras to comprehend the structure of a neutron star as well as  allowing us to test new theories of modified gravity Like: scalar-tensors theories \cite{Langlois:2018dxi,Babichev:2016jom}, scalar-torsion theories \cite{Boumaza:2021fns}, $f(R)$-gravity \cite{kase2019neutron,doneva2020stability} and $f(T)$-gravity \cite{Boehmer:2011gw,Bahamonde:2020snl}. Studying NSs, BHs and the late-time  accelerating universe in the context of  alternative theories of modified gravity is commonly discussed in the literature \cite{Berti:2015itd,Saito:2015fza,kase2019dark,Boumaza:2020klg} because they provide  different  behaviours from general relativity (GR). Munch information, about the mass, radius and tidal deformation of the star, can be carried through the GWs signals which will allow us to rule out or rule in many theories of gravity and EoSs. \\

The tidal love number (TLN) $k_2$ is a constant that links the quadrupole  moment $Q_{ij}$ to the external quadrupolar tidal field $E_{ij}$ through the linear relation $Q_{ij}=(2/3)r_s^5k_2GE_{ij}$ \cite{Hinderer:2007mb}, where $r_s$ is the radius of the star. This constant encodes the information about the internal structure of the NS as well as the behaviour of the gravity in the strong regime and it can be calculated from GWs signals \cite{Flanagan:2007ix}. In other words, the tidal effects are imprinted in the evolution of GWs phase which is caused by the tidal deformation of the NSs. In general relativity, the deformation  of the star as well as its mass and radius, which  depend on the type of the equation of state and the central density, are robust quantities to constraint the equation of state \cite{Ruiz:2017due}, since the  matter at ultra-nuclear densities is not well understood. In addition, in general relativity, we  find  many other tidal coefficients of a self-gravitational body characterized by the integer $l$ and we find also two types of tidal love numbers, called electric type and magnetic type \cite{Damour:2009vw}. In this reference, the author studied in detail the tidal response associated to the electric and magnetic types in the context of general relativity.\\

However, it would be interesting to study the TLNs of neutron star in alternative theories of the gravity where it has been considered in few papers due to the complexity of  its equations. In Ref.\cite{Yagi:2013awa}, the authors took dynamical CS gravity \cite{Alexander:2009tp}, which motivated by superstring theory \cite{Alexander:2004xd} and inflation \cite{Weinberg:2008hq}, as an example to investigate $I$-Love-$Q$ relation in neutron star. For the same objective, the authors, in Ref.\cite{Sham:2013cya}, described NSs in  Eddington-inspired Born-Infeld (EiBI) gravity which is equivalent to GR at low densities and thus the universal relations can be tested in this regime. Furthermore, the effects on  TLNs from the modification of gravity were shown in Ref.\cite{Yazadjiev:2018xxk}, where they found that polar TLN is slightly affected by $R^2$-gravity while the axial TLN might be ten time larger than that in GR. The deviation of generic scalar-tensor theories of gravity, which is equivalent to $f(R)$-gravity, from GR was also investigated by the authors of Ref.\cite{Pani:2014jra}. Inspired by these works, we will consider a theory of gravity with a shift symmetric disformal coupling in which two metrics control the gravitational dynamics and the behavior of the matter field, respectively.\\

Based on Finsler Geometry, disformal transformation was proposed for the first time in Ref.\cite{Bekenstein:1992pj} to generalize the transformation between the two metrics without violating the causality principle. Disformal transformations have been widely suggested not only to explain the origin of the late-time accelerating universe \cite{Zumalacarregui:2012us} but they have been also introduced as an interaction between dark energy and dark matter (DM) \cite{Zumalacarregui:2013pma} to solve the \textit{coincidence problem} \cite{Zlatev:1998tr}. This kind of transformation has many applications in subatomic physics and quantum gravity, for example: it can be a candidate to exotic DM interactions \cite{Bettoni:2011fs,Brax:2020gqg} and it can be also useful to construct a renormalizable theories of gravity \cite{Mukohyama:2013gra}. Additionally, it provides new phenomena called the \textit{disformal screening mechanism} \cite{Koivisto:2012za}, where the scalar field effects is hidden in high density environments. Disformal  transformation is also a useful tool to classify  scalar tensor theories with second order derivatives Like: degenerate-higher-order-scalar-tensor theories (DHOST), for more details see Ref.\cite{BenAchour:2016cay}.\\

Neutron stars with disformal coupling seem to be a good model to understand different physical quantities such as inertia momentum, mass, radius and central density of NS \cite{Minamitsuji:2016hkk} as well as the relation between them. In the later reference, the authors have not studied the proprieties of TLN because it would be very difficult to solve the equations of perturbation analytically  at the exterior of the star, which means also calculating the TLN would be impossible. However, in our paper we will present a model in which the scalar field is vanished outside of the star and thus one can easily show the characteristic of TLN in this model. We will also use the universal relation $C$-$\Lambda$, proposed by  Maselli
et al.\cite{Maselli:2013mva}, to compare our results with the observations of GW170817 event. This relation was utilized in Ref.\cite{Biswas:2019gkw} to analyze the deformation of the compact star with anisotropic pressure and this sounds to be a good approach to discriminate between different EoS.\\

This paper is organized as follow: The first section.\ref{Sec1} is devoted to present the model that we will consider for the next sections and the general equations of the metric and scalar field are also presented for the conformal case. The second section.\ref{Sec2}, the equations of the metric in Einstein  frame is derived in a static and spherically symmetric space time, which allow us to observe the behaviour  of metric and matter at the center of the NS. In Sec.\ref{Sec3}, TLN was examined after calculating the equations of perturbation and considering a particular case.

\section{The model:}\label{Sec1}

{\hskip 2em}In order to study disformal models of gravity, we suppose that the action of our model is described by a \textit{geometrical} (or gravitation) metric $g_{\mu\nu}$ and a \textit{physical} metric $\overline{g}_{\mu\nu}$ (where the matter in the universe follows the geodesics of $\overline{g}_{\mu\nu}$). In other words, $g_{\mu\nu}$ and  $\overline{g}_{\mu\nu}$ are, respectively, the Einstien and Jordan frame metrics and they are related by  the disformal transformation 
\begin{eqnarray}
\overline{g}_{\mu\nu}=\C(X,\varphi)\; g_{\mu\nu}+\D(X,\varphi)\; \varphi_{;\mu}\varphi_{;\nu},\label{disformal transformation}
\end{eqnarray}
where $\C$ (Conformal factor) and $\D$ (Disformal factor)  are arbitrary functions of the scalar field $\varphi$ and the kinetic term $X=\varphi^{;\mu}\varphi_{;\mu}$, where ";" notation represents the covariant derivative. Note that, as long as these functions are positive, the signature $(-,+,+,+)$ is conserved and  the causality is respected \cite{Bekenstein:1992pj}.  The general conformal transformation is recovered by setting $\D=0$. From (\ref{disformal transformation}), it is easy to show that the relations between the inverses and determinants of the two metrics are given by 
\begin{eqnarray}
\overline{g}^{\mu\nu}&=&\C(X,\varphi)^{-1}\; \left(g^{\mu\nu}-\frac{\D(X,\varphi)}{\C(X,\varphi)-2\D(X,\varphi)X)}\; \varphi_{;\mu}\varphi_{;\nu}\right),\label{disformal transformation inverse}\\
\sqrt{-\overline{g}}&=&\sqrt{-g}\;\C(X,\varphi)^2\;\sqrt{1-2\frac{\D(X,\varphi)}{\C(X,\varphi)}X}.
\end{eqnarray}
Also, the transformation of the propagation speed of gravitational  waves between the two frames is given by \cite{Ezquiaga:2017ekz}
\begin{eqnarray}
c_g^2 = \frac{\overline{c}_g^2(X)}{1+2 X \D(X,\varphi)},\label{cg}
\end{eqnarray}
where $c_g$ and $\overline{c}_g$ are propagation speed of GWs in the Einstein frame and in the Jordan frame, respectively. We also consider a general $\FK$ function of  scalar field $\varphi$ and  the term $X$. Supposing so, the action in the Einstein frame can be written as
\begin{eqnarray}
S=\int d^4 x\;\sqrt{-g}\;L_{vac}+\sqrt{-\overline{g}}\; \Lm(\overline{g}_{\mu\nu},\psi),\qquad \text{with}\qquad L_{vac}=\frac{\cR}{2}R+\FK (X,\varphi),\label{S}
\end{eqnarray}
where  $R$ is the Ricci scalar in Einstien frame and  $\cR = (8 \pi G)/c^4$ ($c$ is the speed of light and $G$ is the gravitational coupling in the Einstein frame). $L_m$ represents the Lagrangian density of the matter field, $\psi$, present in the universe. The first  term in action (\ref{S}), called k-essence, is a general form to the action of the quintessence model ($\FK(X,\varphi)= X- V(\varphi)$, where $V(\varphi)$ is the potential), but in this paper we will deal with the shift symmetric models, i.e. the action is invariant  under the transformation $\varphi\rightarrow\varphi + const$. Thus, the functions $\FK$, $\C$ and $\D$ will depend only on the kinetic term $X$. In addition, because of the constraint imposed by the detection of gravitational waves due to a collision of BNs (GW170817) \cite{LIGOScientific:2017vwq} followed by a short gamma ray burst \cite{Goldstein_2017}, the speed of gravitational waves in scalar tensor theories frame work  are reduced in order to obtain the speed of light \cite{Ezquiaga:2017ekz}. Thus, according to Eq.(\ref{cg}), we impose  $\D = 0$ to respect this constraint of GW170817 event because $c_g=1$ for the k-essence models. We can also impose a special function of  the disformal factor $\D$ to obtain $\overline{c}_g=c$ but this issue will not discussed in this paper. Therefore, the functions of our model are reduced to
\begin{eqnarray}
\D = 0, \qquad \FK = \FK(X)\qquad and \qquad \C = \C(X).\label{functions}
\end{eqnarray}
In this case, the action become invariant under the transformation $\varphi\rightarrow-\varphi$ and it is look like  beyond Horndeski with two metrics
\begin{eqnarray}
\frac{2}{\cR}L_{vac}=\frac{1}{\C}\overline{R}+\frac{3}{2}\C_X^2 \varphi_{;\mu}\varphi_{;\mu\alpha}\varphi^{;\alpha\nu}\varphi_{;\nu}+\frac{2}{\cR \C^2}\FK,
\end{eqnarray}   
where $\overline{R}$ is the Ricci scalar in Jordan frame. Since $X$ is a complex function of $g_{\mu\nu}$, we can not derive the full expressions of $g_{\mu\nu}$ from Eq.(\ref{disformal transformation}) with the functions (\ref{functions}). Now, we define the contravariant energy-momentum tensor in Jordan frame as
\begin{eqnarray}
\overline{T}^{\mu\nu(m)}=\frac{2}{\sqrt{-\overline{g}}}\frac{\delta(\sqrt{-\overline{g}}L_m)}{\delta\overline{g}_{\mu\nu}}=(\overline{\rho}+\overline{P})\overline{u}^\mu\overline{u}^\nu +\overline{g}^{\mu\nu}\overline{P},\label{Tmunu}
\end{eqnarray}
where $\overline{P}$ and $\overline{\rho}$ are the pressure and the energy density of the fluid in Jordane frame. $\overline{u}^\mu$ is the four velocity of the fluid and it is constrained by $\overline{u}^\mu\overline{u}_\mu=-1$. Provided that the matter action is invariant under the coordinate transformation $x^\mu\rightarrow x^\mu+ \xi^\mu$, the energy-momentum tensor is covariantly conserved \cite{Zumalacarregui:2013pma}. So, the conservation equation of the fluid in Jordan frame reads
\begin{eqnarray}
\overline{\nabla}_{\mu}\overline{T}^{\mu\nu(m)}=0.\label{dtmunu}
\end{eqnarray}
We note that the covariant derivative is  constructed using $\overline{g}_{\mu\nu}$ and its derivatives. In Einstien frame, the tensor energy-momentum is given by 
\begin{eqnarray}
T^{\mu\nu(m)}=\frac{2}{\sqrt{-g}}\frac{\delta(\sqrt{-\overline{g}}L_m)}{\delta g_{\mu\nu}}=(\rho + P)u^\mu u^\nu +g^{\mu\nu}P,\label{TmunuE}
\end{eqnarray}
where $\rho$, $P$ and $u^\mu$ are the energy density, the pressure and the four velocity of the matter present in the universe.  The relation between the tensors $T^{\mu\nu(m)}$ and $\overline{T}^{\mu\nu(m)}$ reads
\begin{eqnarray}
T^{\mu\nu(m)}=\sqrt{\frac{\overline{g}}{g}}\frac{\delta \overline{g}_{\mu\nu}}{\delta g_{\mu\nu}}\overline{T}^{\mu\nu(m)}=\C^3 \overline{T}^{\mu\nu(m)},\label{TmunuE}
\end{eqnarray}
Unlike in the Jordan frame the tensor $T^{\mu\nu(m)}$ is not conserved $\nabla_{\mu}T^{\mu\nu(m)}\neq 0$ in Einstien frame. This can be shown by taking the covariant derivative $\overline{\nabla}_{\mu}$ of the equation (\ref{TmunuE}) and by using the relations in Appendix. A of Ref.\cite{Zumalacarregui:2013pma}. Doing so, we obtain
\begin{eqnarray}
\nabla_\mu T^{\mu\nu(m)}=\frac{\C_X}{2\C}\left(T^{(m)}\left(\partial^\nu X-\varphi_{;\alpha}\varphi^{;\alpha\nu}\right)+6\varphi^{;\alpha}\varphi_{;\alpha\beta}T^{\beta\nu(m)}\right),\label{dTmunuE}
\end{eqnarray}
where $T^{(m)}$ is the trace of the tensor $T^{\mu\nu(m)}$. However, the total energy-momentum tensor of the model is covariantly conserved with respect to $g_{\mu\nu}$, which is due to Bianchi identities. The general equations of metric and scalar fields are derived by using the Euler-Lagrange equations
\begin{eqnarray}
&\partial_{\alpha}\partial_{\beta}\left[\frac{\partial\sqrt{-g} L_{tot}}{\partial(\partial_{\alpha}\partial_{\beta}g_{\mu\nu})}\right]-\partial_{\alpha}\left[\frac{\partial\sqrt{-g}L_{tot}}{\partial(\partial_{\alpha}g_{\mu\nu})}\right]+\frac{\partial\sqrt{-g} L_{tot}}{\partial g_{\mu\nu}}=0&,\label{e1}\\
&\nabla_{\mu} J^\mu = 0,&\label{ephi}
\end{eqnarray}
where
\begin{eqnarray}
L_{tot}&=& L_{vac}+\sqrt{\frac{\overline{g}}{g}}\Lm(\overline{g}_{\mu\nu},\psi),\label{Ltot}\\
J^\mu &=&  \frac{\partial L_{tot}}{\partial(\varphi_{;\mu})}.\label{J}
\end{eqnarray}
The equations (\ref{e1}) and (\ref{J}) can be simplified, by using the equations (\ref{Ltot}) and (\ref{TmunuE}), to
\begin{eqnarray}
&\partial_{\alpha}\partial_{\beta}\left[\frac{\partial\sqrt{-g} L_{vac}}{\partial(\partial_{\alpha}\partial_{\beta}g_{\mu\nu})}\right]-\partial_{\alpha}\left[\frac{\partial\sqrt{-g}L_{vac}}{\partial(\partial_{\alpha}g_{\mu\nu})}\right]+\frac{\partial\sqrt{-g} L_{vac}}{\partial g_{\mu\nu}}=-\frac{\sqrt{-\overline{g}}}{2}\overline{T}^{\alpha\beta(m)}\frac{\partial\overline{g}_{\alpha\beta}}{\partial g_{\mu\nu}}&,\label{e12}\\
& J^\mu= \frac{\partial L_{vac}}{\partial(\varphi_{;\mu})}- \frac{\sqrt{-\overline{g}}}{2}\overline{T}^{\alpha\beta(m)}\frac{\partial\overline{g}_{\alpha\beta}}{\partial\varphi_{;\mu}} \label{J2}&
\end{eqnarray}
These equations remain true for the general disformal transformation (\ref{disformal transformation}) but if $L_{vac}$ depends on the second derivatives of the scalar field $\varphi_{;\mu\nu}$, the equation (\ref{J2}) must be expanded by adding the term $\partial_\nu(\partial L_{vac}/ \partial(\varphi_{;\mu\nu}))$ to Eq.(\ref{J2}). Equations  (\ref{ephi}), (\ref{e12}) and (\ref{J2}) will be used to derive the equations of the metric and scalar field at the background and perturbed levels.

\section{Background equations}\label{Sec2}
In this section we wish to study the behaviour of the model described by the action (\ref{S}) and the functions (\ref{functions}) in a static and spherically symmetric spacetime, endowed with the metric
\begin{eqnarray}
ds^2 = -f(r)^2\, dt^2+ h(r)^2\, dr^2+ q(r)^2 r^2\,(d\theta^2 +\sin^2 \theta \, d\phi^2),\label{dsE}
\end{eqnarray}
where $f(r)$, $h(r)$ and $q(r)$ are functions of radial coordinate $r$ in the Einstein frame. As a consequence of the spacetime symmetries, we assume that the scalar field is spherically symmetric, i.e. the scalar field depends only on $r$ ($\varphi=\varphi(r)$), and thus the functions $\FK$ and $\C$ depend also only on $r$. From the transformation (\ref{disformal transformation}), the metric in Jordan frame reads
\begin{eqnarray}
ds^2 = \frac{1}{\C}\left(-\overline{f}(r)^2\, dt^2+ \overline{h}(r)^2\, dr^2+ \overline{q}(r)^2 r^2\,(d\theta^2 +\sin^2 \theta \, d\phi^2)\right).\label{dsJ}
\end{eqnarray}
Since, the metric in Jordan frame is  complicated due to coefficient $1/\C$ which contains the metric $h(r)$, it is convenient to study the NSs  in Einstien frame, but we will use the pressure, the energy density and the four-velocity vector of the fluid  in Jordan frame, to derive the equations of motion. From the constraint $\overline{g}_{\mu\nu}\overline{u}^\mu\overline{u}^\nu=-1$ with $\overline{u}^\mu=(\overline{u}^t,0,0,0)$ in a static and spherically symmetric spacetime, we have $\overline{u}^t=\frac{1}{f(r)\sqrt{\C}}$. Thus, the tensor energy momentum in the two frames are written as function of the metric (\ref{dsE}) as
\begin{eqnarray}
T^{\mu\nu(m)}=\C\; \overline{T}^{\mu\nu(m)}=Diag\left(\frac{\overline{\rho}}{f(r)^2},\frac{\overline{P}}{h(r)^2},\frac{\overline{P}}{r^2 q(r)^2},\frac{\overline{P}}{r^2\sin^2 \theta q(r)^2}\right).
\end{eqnarray} 
where $\overline{\rho}$ and $\overline{P}$ are only a function of $r$. Unlike in the disformal transformation ($D\neq 0$) case in which the matter becomes anisotropic ($P_r\neq \overline{P}_t$ where $P_r$ is the radial pressure and $\overline{P}_t$ is the tangential pressure of the matter) in the Einstein frame, the fluid is always isotropic. The pressure of a neutron star is related to the energy density through the  different forms of the equations of state, but we prefer the ones, simulated from binary neutron star coalescence, defined by

\begin{eqnarray}
\zeta &=& \frac{\text{a}_1+\text{a}_2 \xi +\text{a}_3 \xi ^3}{(\text{a}_4 \xi +1)}d\left(\text{a}_5 (\xi -\text{a}_6)\right)+(\text{a}_7+\text{a}_8 \xi)d\left(\text{a}_9 (\text{a}_{10}-\xi )\right)+(\text{a}_{11}+\text{a}_{12} \xi)d\left(\text{a}_{13} (\text{a}_{14}-\xi )\right)\nonumber\\
& & +(\text{a}_{15}+\text{a}_{16} \xi )d\left(\text{a}_{17} (\text{a}_{18}-\xi )\right)
 +\frac{\text{a}_{19}}{\text{a}^2_{20} (\text{a}_{21}-\xi )^2+1}+\frac{\text{a}_{22}}{\text{a}^2_{23} (\text{a}_{24}-\xi )^2+1},\label{EOS}
\end{eqnarray} 
with
\begin{eqnarray}
d(x)=\frac{1}{e^x+1}.
\end{eqnarray}
where we have utilized the analytic expressions of the EoS  for  realistic NSs  of Refs.\cite{Haensel:2004nu,Potekhin:2013qqa}. In these Refs, they  introduce the variables $\xi=log_{10}(\overline{\rho}/g\;cm^{-3})$ and $\zeta=log_{10}(\overline{P}/g\;cm^{-3})$ to express the parametrization of $\overline{P}(\overline{\rho})$. The  parameters $\text{a}_i$ can be found in Ref.\cite{Haensel:2004nu} for  the  SLy  and  FPS  and in \cite{Potekhin:2013qqa} for  the  BSk19, BSk20 and BSk21. However, the coefficients of our  EoS's parametrezation is a bit different from those in these literature like: for SLy  and  FPS, we have $\text{a}_{19.....24}=0$ and for BSk19, BSk20 and BSk21 the notations in \cite{Potekhin:2013qqa} are found by performing the transformations $\text{a}_{10}\rightarrow\text{a}_6$ and $\text{a}_{11....24}\rightarrow\text{a}_{10....23}$ \cite{Kase:2020yhw}.  \\

If we substitute the metric (\ref{dsE}) in $L_{vac}$, we find that the corresponding Lagrangian can be written as function of $f(r)$, $h(r)$, $q(r)$ and $\varphi(r)$ as well as its derivative. Doing so and after integrating by part to eliminate the second derivatives of the metrics $f(r)$ and $q(r)$, it follows that
\begin{eqnarray}
&L_{vac}=\cR  \left( \frac{q'(r)^2}{h(r)^2 q(r)^2}+\frac{2 f'(r)q'(r)}{f(r) h(r)^2 q(r)}-\frac{2 q'(r)}{r h(r)^2 q(r)}+\frac{2 h'(r)}{r h(r)^3}+\frac{1}{r^2q(r)^2}-\frac{1}{r^2h(r)^2}\right)+\FK , &\label{Lvac backgound}
\end{eqnarray}
where the prime denotation "$'$" corresponds to the radial derivative. Substituting the above Lagrangian in the equations (\ref{e12}) and taking the limits $q(r)\rightarrow 1$, $q'(r)\rightarrow 0$ and $q''(r)\rightarrow 0$, the components $(t,t)$, $(r,r)$ and $(\theta,\theta)$ read, respectively
\begin{eqnarray}
\frac{h'(r)}{h(r)}&= & \frac{1-h(r)^2}{2 r}-\frac{r h(r)^2 \left(\FK-\C^2 \overline{\rho}(r)\right)}{2 \cR},\label{eh}\\
\frac{f'(r)}{f(r)}&= & \frac{r  \left(\varphi'(r)^2 \left(\C \C_{X}(\overline{\rho}(r)-3  \overline{P}(r))-2 \FK_{X}\right)+h(r)^2 \left(\C^2 \overline{P}(r)+\FK\right)\right)}{2 \cR}\nonumber\\
&&+\frac{ \left(h(r)^2-1\right)}{2 r},\label{ef}\\
\frac{f''(r)}{f(r)}&= &\frac{h(r)^2 \left(\C^2 \overline{P}(r)+\FK\right)}{\cR }+\frac{h'(r)}{r h(r)}+\frac{f'(r)}{f(r)}\left(\frac{h'(r)}{h(r)}-\frac{1}{r}\right).\label{eff}
\end{eqnarray}
When $\C_{X}= 0$ the equations of theory are equivalent to those in  GR. Similarly, from Eq. (\ref{dtmunu}), the equation corresponding to the matter of the stellar object in the metric (\ref{dsE}) is found as 
\begin{eqnarray}
\frac{\overline{P}'(r)}{\overline{P}(r)+\overline{\rho}(r)}&=& \frac{\C_{X}}{\C h(r)^2}\left(\frac{h'(r) }{h(r)} \varphi '(r)^2-\varphi '(r) \varphi ''(r)\right)-\frac{f'(r) }{f(r)}.\label{eP}
\end{eqnarray}
The terms $f'(r)$ and $h'(r)$ can be  replaced by using the Eqs. (\ref{eh}) and (\ref{ef}). The obtained equation contains  the scalar field and its derivatives which are eliminated by solving the equation
\begin{eqnarray}
\frac{d}{dr}\left[f(r)h(r)r^2 J^r\right]&=&0,\label{ephi2}
\end{eqnarray}
with
\begin{eqnarray}
J^r=\frac{\varphi '(r) }{h(r)^2}\left(\C \C_{X}(\overline{\rho}(r)-3  \overline{P}(r))+2 \FK_{X}\right).\label{eJ}
\end{eqnarray}
The equation (\ref{eff}) is a combination of Eqs. (\ref{eh}), (\ref{ef}), (\ref{eP}) and  (\ref{ephi2}) and thus this equation will be neglected in our work. In addition, one can reduce our system of equations by considering the  particular solution $J^r=0$ which has the solutions: $\varphi '(r)=0$ (This case is equivalent to General relativity) and 
\begin{eqnarray}
\C \C_{X}(\overline{\rho}(r)-3  \overline{P}(r))+2 \FK_{X}=0,\label{eX}
\end{eqnarray}
when $\C_{X}\neq 0$. From this equation, the kinetic term $X$ can be written as a function of the quantity $\overline{\rho}(r)-3  \overline{P}(r)$ for a special form of functions $\C$ and $\FK$. In order to have a good description of the NS, the  forms of this functions must be chosen to avoid singularity from the surface of the star $r_s$ to the center of the star, example:  for $\FK_{X}= a X$ and $\C=1+b X$, we have $X\sim (2a-b(\overline{\rho}(r)-3  \overline{P}(r)))/(\overline{\rho}(r)-3  \overline{P}(r))$, and it follows that  the singularity can not be avoided at the surface of the star. After solving Eq.(\ref{eX}) for $\varphi'(r)$ and substituting the result in Eqs. (\ref{eh}), (\ref{ef})  and (\ref{eP}),  we arrive to the final equations  associated to $h(r)$, $f(r)$ and $\overline{P}(r)$. Note that the equations (\ref{eh}) and (\ref{eP}) are independent from $f(r)$ and its derivatives and thus Eq.(\ref{ef}) can be solved after obtaining the numerical or analytical solution  of the quantities $h(r)$ and $\overline{P}(r)$.\\

Now, the irregularity at the center ($r=0$) of NSs might be avoided, if we must impose the boundary conditions: $\overline{P}'(0)=\overline{\rho}'(0)=f'(0)=h'(0)$. To show this, let's express the solutions  around $r=0$, as
\begin{eqnarray}
&& \overline{P}(r)=\overline{P}_c+\sum^\infty_{i=2}\overline{P}_i r^i, \qquad \rho(r)=\rho_c+\sum^\infty_{i=2} \rho_i r^i,\qquad h(r)=1+\sum^\infty_{i=2} h_i r^i,\nonumber\\
&& f(r)=f_c+\sum^\infty_{i=2} f_i r^i,\label{series}
\end{eqnarray}
where $P_c$, $\rho_c$, $f_c$,  $\overline{P}_i$, $\overline{\rho}_i$, $f_i$ and $h_i$ are constants and the subscript $c$ means the value of each function at the center of the NS. In this development, we have  chosen the condition $h(0)=1$ to have regular solutions at the center of the star and the constant $\overline{\rho}_c$, which is the central energy density of the NS, is related to the pressure $\overline{P}_c$ through the equation of state $\overline{P}_c(\overline{\rho}_c)$. In this case, due to the constraint (\ref{eX}) and after choosing the functions  the series development of the scalar field is
\begin{eqnarray}
\varphi'(r)=\varphi_c' +\sum^\infty_{i=2} \varphi_i' r^i, \label{sereisphi}
\end{eqnarray}
where $\varphi_c'$ and $\varphi_i'$ are real constants and they can be written as function of $\overline{\rho}_c$, $\overline{P}_c$, $\overline{\rho}_2$, $\overline{P}_2$ and the parameter of the model (we will see this in the next, for a spacial case). From the series (\ref{sereisphi}), we have $\varphi''(r=0)=0$ instead of $\varphi'(r=0)=0$ and since the model is shift symmetric, the regularity is always satisfied as well as we set $\varphi(r=0)=0$ without loss of generality. Inserting Eq. (\ref{series}) in Eqs. (\ref{eh}), (\ref{ef})  and (\ref{eP}) and solving the equations proportional to $r$ for $f_2$, $h_2$ and $P_2$, the behaviour of the pressure and the metrics are
\begin{eqnarray}
\overline{P}&=&\overline{P}_c+\overline{P}_2 r^2+ O\left(r^3\right),\label{idPJ}\\
h&=&1+\frac{\C^2 \overline{\rho}_c-\FK}{6 \cR }r^2+ O\left(r^3\right),\label{idhE}\\
\frac{f}{f_c}&=&1+ \frac{ 3 \left(\varphi _c'\right)^2 \left(\C \C_X \left(\overline{\rho}_c-3 \overline{P}_c\right)-2 \FK_X\right)+2 \FK+\C^2\left(3  \overline{P}_c+ \overline{\rho}_c\right)}{12 \cR} r^2+ O\left(r^3\right),\label{idfE}
\end{eqnarray}
where functions $\FK$ and $\C$ depend on $X_c=\left(\varphi _c'\right)^2$ and $\overline{P}_2$ is obtained by solving the equation
\begin{eqnarray}
\overline{P}_2&=&-\frac{\overline{P}_c+\overline{\rho}_c}{12 \C  \cR} \left(\left(\varphi_c'\right){}^2 \left(\C^2 \C_X \left(\overline{\rho}_c-9 \overline{P}_c\right)-6 \C \FK_X +2 \C_X \FK\right)\right.\nonumber\\
&&\left.+\C^3 \left(3 \overline{P}_c+\overline{\rho}_c\right)+2 \C \FK+12 \C_X \cR \varphi _2' \varphi _c'\right)
\end{eqnarray}  
Because of the conformal function, the deviation of the metric and the pressure are expected to deviate from those in GR far from the center of the star. The metrics $\overline{f}$ and $\overline{h}$, in Jordan frame, behave as
\begin{eqnarray}
\frac{\overline{f}}{f_c\C^{\frac{1}{2}}}&=&1+\frac{1}{12 \cR } \left(\frac{2 \C_X }{\C}\varphi _c' \left(\varphi _c' \left(\FK-\C^2 \overline{\rho}_c\right)+6 \cR  \varphi _2'\right)+\C^2 \left(3 \overline{P}_c+ \overline{\rho}_c\right)\right.\nonumber\\
&&\left. +\varphi _c'^2 \left(\C \C_X \left(3 \overline{\rho}_c-9 \overline{P}_c\right)-6 \FK_X\right)+2 \FK\right)r^2 + O\left(r^3\right),\label{idfJ}\\
\frac{\overline{h}}{\C^{\frac{1}{2}}}&=&1+\frac{1}{6 \cR \C} \left(\C_X \varphi_c' \left(\varphi_c' \left(\FK-\C^2 \overline{\rho}_c\right)+6 \cR  \varphi_2'\right)+\C \left(\C^2 \overline{\rho }_c-\FK\right)\right)r^2 + O\left(r^3\right).\label{idhJ}
\end{eqnarray}
From the relation (\ref{TmunuE}), the pressure of the matter around the center is written as
\begin{eqnarray}
\frac{P}{\C^2}&=& \overline{P}_c+\left(\overline{P}_2+\frac{1}{3 \C \cR } \left(2 \C_X \overline{P}_c \varphi_c' \left(\varphi_c' \left(\FK - \C^2 \overline{\rho }_c\right)+6 \cR  \varphi _2'\right)\right)\right)r^2+ O\left(r^3\right).\label{idhPE}
\end{eqnarray}
The constant $\overline{P}_2$ correspond to term proportional to $r^2$ in Eq.(\ref{idPJ}). In the two frames, the metrics and the pressure of the fluid inside the star do not contain singularity at the center of the star as well as they have a different comportment from those in GR as long as $\C\neq 1$. Note that in the next we we use the notation $f$, $h$.....etc instead of $f(r)$, $h(r)$.....etc.

\section{Tidal love Number of neutron star}\label{Sec3}

{\hskip 2em}In this section, we will construct the perturbed equations around the background,  by following the same steps presented in the last sections. First, we expand the metric, in the Einstein frame, as:  $g_{\mu\nu}=g_{\mu\nu}^{(0)}+g_{\mu\nu}^{(1)}$ where $g_{\mu\nu}^{(0)}$ corresponds to the background (\ref{dsE}) and $g_{\mu\nu}^{(1)}$ is the perturbed metric. Second,  we consider that $g_{\mu\nu}^{(1)}$ is a function of the coordinates $(r,\theta,\phi)$ and it is  decomposed into polar and axial mode as:
\begin{eqnarray}
g_{\mu\nu}^{(1)}=h_{\mu\nu }^{(E)}+h_{\mu\nu}^{(B)},
\end{eqnarray} 
 where $h_{\mu\nu }^{(E)}$ (polar mode) and $h_{\mu\nu}^{(B)}$ (axial mode) have, respectively, the parities $(-1)^l$ and $(-1)^{l+1}$  under the rotation in the two dimensional plane $(\theta,\phi)$. By performing this transformation, the components $g_{t,t}$, $g_{r,t}$ and $g_{r,r}$ transform as scalars, the components $g_{t, \theta}$, $g_{t,\phi}$, $g_{r,\theta}$ and $g_{r,\phi}$ transform as vectors as well as  the components $g_{a, b}$, with $(a,b)=(\theta,\phi)$ is considered as components of a tensor. To derive the equations of these competents, we start by defining the tensors $h_{\mu\nu }^{(E)}$ and $h_{\mu\nu}^{(B)}$ in the Einstein frame as 
\begin{eqnarray}
ds^{(E)}&=&\sum^{lm}\left(f^2 H_{0}^{(lm)}dt^2+2 H_{1}^{(lm)}dtdr+h^2 H_{2}^{(lm)}dr^2+H_{3}^{(lm)}r^2 d\theta^2+ H_{4}^{(lm)}r^2\sin^2\theta d\phi^2\right)\nonumber\\
&&\times Y_{lm}+2H_{5}^{(lm)}h^2 (\partial_\theta Y_{lm}drd\theta+\sin\theta\partial_\phi Y_{lm}drd\phi ),\label{dsEE}\\
ds^{(B)}&=&2\sum^{lm}\left( h_{0}^{(lm)} dtd\phi+  h_{1}^{(lm)} drd\phi\right)\sin\theta\partial_\theta Y_{lm}-\left( h_{0}^{(lm)} dtd\theta+  h_{1}^{(lm)} drd\theta\right)\frac{\partial_\phi Y_{lm}}{\sin\theta},\label{dsEB}
\end{eqnarray}
where  $Y_{lm}(\theta,\phi)$ are the spherical harmonics and the metrics $H_i^{(lm)} $ and  $h_i^{(lm)} $ are radial functions. Note that after deriving the equations of the even modes, we set  $H_{5}^{(lm)} =0$ and $H_{3}^{(lm)} =H_{4}^{(lm)}$, which correspond to the metric in the Regger-Wheeler gauge.  This formulation will allow us to study the even and odd modes separately which is due to the decoupling between the odd and even metrics at the linearized level. Without loss of generality, we will work with $m=0$ which means the metric will depend only on $\theta$. We will also work with notations $H_{i}$ and $h_{i}$  instead of  $H_{i}^{(lm)}$ and $h_{i}^{(lm)}$.

\subsection{Axial:}

{\hskip 2em}Since the scalar field, pressure and energy density are scalar quantities, the perturbation associated to them are vanished. Thus, the second order of the Lagrangian perturbation in the axial mode reads
\begin{eqnarray}
S^{(2)}_{vac,\, axial}&=&\sum^{l}\int dr l(l+1)\left(\frac{\cR }{4 f h}h_0'^2+\frac{ \cR  \left(f \left(h^3 l+2 r h'\right)-2 h r f'\right)+2 f \FK h^3 r^2}{4 f^2 h^2 r^2}h_0^2\right.\nonumber\\
&&\left.+ \frac{\cR \left(4 r f'+f \left(2-h^2 l\right)\right)+f h_1^2\left(4 r^2 \FK_X X-2 \FK  r^2\right)}{4 h^3 r^2} h_1^2\right).\label{S2odd}
\end{eqnarray}
Substituting this action in (\ref{e1}), we have $h_1=0$ and 
\begin{eqnarray}
h_0''=\left(\frac{f'}{f}+\frac{h'}{h}\right)h_0'+\left(\frac{2h'}{hr}-\frac{2f'}{fr}+\frac{2 h^2}{\cR }\left(\C^2 \overline{P}+\FK\right)+\frac{h^2l(l+1)}{r^2}\right)h_0.
\end{eqnarray}
In contrast to GR, this equation does not have analytical solution in the vacuum due to the $\FK$ term.

\subsection{Polar:}

{\hskip 2em}As we have done to the metric, we also perturb the scalar field, the pressure and the energy density, respectively, as:
\begin{eqnarray}
\varphi = \varphi(r)+\sum^{l}Y_{l0}\delta\varphi, \qquad\overline{P}=\overline{P}(r)+\sum^{l}Y_{l0}\delta \overline{P}, \qquad\overline{\rho}=\overline{\rho}(r)+\sum^{l} Y_{l0}\delta \overline{\rho}.
\end{eqnarray}
Note that $\delta\varphi$, $\delta \overline{P}$ and $\delta \overline{\rho}$ are also functions of $r$. Substituting the metric (\ref{dsEE}) in the action $(S)^{(2)}_{vac,\, polar}=\int d^4 x \sqrt{-g}L_{vac,\, polar}$ and after integrating by parts, the second order of even mode perturbations are expressed as
\begin{eqnarray}
S^{(2)}_{vac,\, polar}&=&\sum^{l}\int dr\left(\overline{P}H_0^2+H_0\left(p_1(H_4''+H_3'')+p_2 H_2'+p_3(H_4'+H_3')+p_4 H_5'+p_5\delta\varphi' \right.\right.\nonumber\\
&&\left.\left. +p_6 (H_4+H_3)+p_7 H_2+p_8 H_3+p_9 H_5\right)+p_{10} H_2^2+H_2\left(p_{11}(H_4'+H_3')+p_{12}\delta\varphi' \right.\right.\nonumber\\
&&\left.\left. +p_{13} (H_4+H_3)+p_{14} H_3+p_{15} H_5\right)+p_{16} H_5^2+H_5\left(p_{17}(H_4'+H_3') +p_{18}H_3'+p_{19}\delta\varphi\right)\right.\nonumber\\
&&\left. +p_{20}H_1^2+p_{21}(H_3+H_4)^2+p_{22} (H_3+H_4)\delta\varphi' +p_{23}\delta\varphi^2 +p_{24}H_3'H_4'+p_{25}\delta\varphi'^2 \right),\nonumber\\\label{S2even}
\end{eqnarray}
where the coefficients $a_{i}$ with $i=\{0,1,2,....,25\}$ are given in the appendix (\ref{A1}). We vary the full action $S^{(2)}_{vac}+S^{(2)}_{m}$ with respect to the perturbed metric and the perturbed scalar field and by using Eq.(\ref{e1}) and  (\ref{ephi}) to obtain different constraints and equations, then the obtained result is simplified by taking the limit $H_{5}=0$ and $H_{3}=H_{4}$. Then, we eliminate $\delta\overline{P}$ and $\delta\overline{\rho}$ by using Eqs. (\ref{deltaP}) and (\ref{deltarho}). The first constraint is $H_1=0$, since there is no radial derivative of this metric in the full action. The second one obtained, by varying with respect $S^{(2)}_{vac,\,polar}+S^{(2)}_{m}$ to $H_3$ and $H_4$, as
\begin{eqnarray}
E_{H_3}&=&E_{H_4}+\frac{h^2(1+l(l+1))}{r^2}(H_0-H_2).\label{eH3}
\end{eqnarray} 
Since $E_{H_3}=0$ and $E_{H_4}=0$, it follows that $H_0=H_2=H$ and
\begin{eqnarray}
E_{H_3}&=& q_0(H''-H_4'')+q_1 H_4'+q_2 H'+q_3 \delta \varphi '+q_4 H+q_5 H_4\label{eH4}
\end{eqnarray} 
where $q_i$  are also written as a function of the background metric and non perturbed scalar field (see Appendix.\ref{A3}). One can eliminate $H_4$ and it derivatives by using the equation associated to $H_5$, $H_2$ and $H_0$, which are given by 
\begin{eqnarray}
E_{H_5}&=&H_4'-H'-\frac{2 f'}{f}H+\frac{4 \FK_X \varphi'}{\cR }\delta \varphi, \label{eH5}\\
E_{H_2}&=& e_0  H_4'+e_1 H'+e_2 \delta \varphi' +e_3 H+e_4 H_4,\label{eH2}\\
E_{H_0}&=& e_5 H_4''+e_6 H_4'+e_7 H'+e_8\delta\varphi'+ e_9 H+e_{10} H_4 ,\label{eH0}
\end{eqnarray}
Finally, the equation of scalar field perturbation can be written in the form 
\begin{eqnarray}
E_{\delta\varphi} &=&\frac{d}{dr}\left[q_6\delta\varphi' +q_7 H +q_8H_4 \right]-2p_{23}\delta\varphi,\label{edelaphi}
\end{eqnarray}
The final equations are derived by eliminating  $H_4$ and its derivatives in Eqs.(\ref{eH3}) and (\ref{edelaphi}) by using Eq. (\ref{eH5}), (\ref{eH2}) and  (\ref{eH0}). Doing so, we find two equation with the forms
\begin{eqnarray}
H''+Y_0H'+Y_1\delta\varphi' +Y_2 H+Y_3\delta\varphi=0,\label{eH}\\
\delta\varphi'' + Z_0 H'+Z_1 \delta\varphi' +Z_2 H +Z_3\delta\varphi=0,\label{eHphi}
\end{eqnarray}
where the coefficients $Y_i$ and $Z_i$ with $i=\{1,2,3\}$ are written as a function of the background metric, pressure and the energy density, see Appendix \ref{A3}. But, we can not have the analytical solution at the exterior of the star, because their expressions in the general form is very long. 

\subsection{Asymptotic behaviour at the interior and exterior of the star}

{\hskip 2em}To have simplified equations and show the central and external behaviour of the metric perturbation, we will take special functions of $\C$ and $\FK$ as
\begin{eqnarray}
\C= 1+ \cC X^{\frac{1}{2}}\qquad\text{and}\qquad \FK= \cK X,\label{spetialfuctions}
\end{eqnarray}
where $\cC$ and $\cK$ are real constants. Solving  the Eq.(\ref{eX}) for $\varphi'$, gives
\begin{eqnarray}
\varphi'=\frac{\cC h (3 \overline{P}-\overline{\rho} )}{4 \cK -\cC^2 (3 \overline{P} -\overline{\rho})}.\label{edphi}
\end{eqnarray}
In this particular model the scalar field is equal to zero outside the star which means that when $r>r_s$, the background and the odd perturbation solutions in the two frames are identical ($\overline{g}_{\mu\nu}^{(0)}=g_{\mu\nu}^{(0)}$ and $\overline{h}_{\mu\nu}^{(B)}=h_{\mu\nu}^{(B)}$) and they are equivalent to those in GR:
\begin{eqnarray}
\overline{f}=\frac{1}{\overline{h}}=f=\frac{1}{h}=\left(1-\frac{2 M}{r}\right)^{\frac{1}{2}},&\\
\overline{h}_0=h_0=c_1 \psi_l^p(r)+c_2 \psi_l^q(r),
\end{eqnarray}
where $c_1$ and $c_2$ are constants of integration and $M$ is the mass of NS. The function $\psi_l^p$ is a growing type solution where $\psi_l^p(r\rightarrow\infty)\approx r^{l+1}$ while $\psi_l^p$, called decreasing type, behaves as $r^{-l}$ when $r$ tends to $\infty$. As example: for $l=2$ the full expression of these functions are
\begin{eqnarray}
\psi_l^q &=&\frac{ 2 M \left(2 M^2 r+2 M^3+3 M r^2-3 r^3\right)+3 r^3 (r-2 M) \log \left[r/(r-2 M)\right]}{24 M^5 r},\\
\psi_l^p &=&  r^2 (r-2 M).
\end{eqnarray}
However, the Einstein  and Jordan frames are not identical for the polar type  since $\overline{H}_0=H-\cC \delta\varphi'/h$ and $\delta\varphi'\neq 0$. In the vacuum, Eqs. (\ref{eH}) and (\ref{eHphi}) are reduced to 
\begin{eqnarray}
H''+\frac{2 M-2 r}{2 M r-r^2}H'+\frac{2 (l+1) l M r-l (l+1) r^2-4 M^2}{r^2 (r-2 M)^2}H=0,\label{eHGR}\\
\delta\varphi''+\frac{2 M-2 r }{2 M r-r^2}\delta\varphi '+\frac{(l+1) l }{2 M r-r^2}\delta\varphi = 0.
\end{eqnarray}
We note that these equations are decoupled and Eq.(\ref{eHGR}) is the same as what we found in GR. These two equations have analytical solutions similar to the odd perturbation solution and they are given by \cite{Damour:2009vw}
\begin{eqnarray}
H=c_1 P_l^2\left[\frac{r}{M}-1\right]+c_2 Q_l^2\left[\frac{r}{M}-1\right],\\
\delta\varphi =\overline{c}_1 P_l^0\left[\frac{r}{M}-1\right]+\overline{c}_2 Q_l^0\left[\frac{r}{M}-1\right],
\end{eqnarray}
where $\overline{c}_1$ and $\overline{c}_2$ are also constants of integration. $P_l^m$ and $Q_l^m$  are the  associated Legendre functions of first and second kind, respectively. The asymptotic behaviour of $H$, $\delta\varphi$ and $\overline{H}_0$ at infinity is as follow
\begin{eqnarray}
H= c_1 r^2+ \frac{c_2}{r^3},\quad \delta\varphi= \overline{c}_1 r^2+ \frac{\overline{c}_2}{r^3}\quad\text{and}\quad \overline{H}_0=c_1 r^2+2 \cC \overline{c}_1 r+ \frac{c_2}{r^3}-3\cC\frac{\overline{c}_2}{r^4}.
\end{eqnarray}
\begin{figure}[h]
\centering 
\includegraphics[scale=0.8]{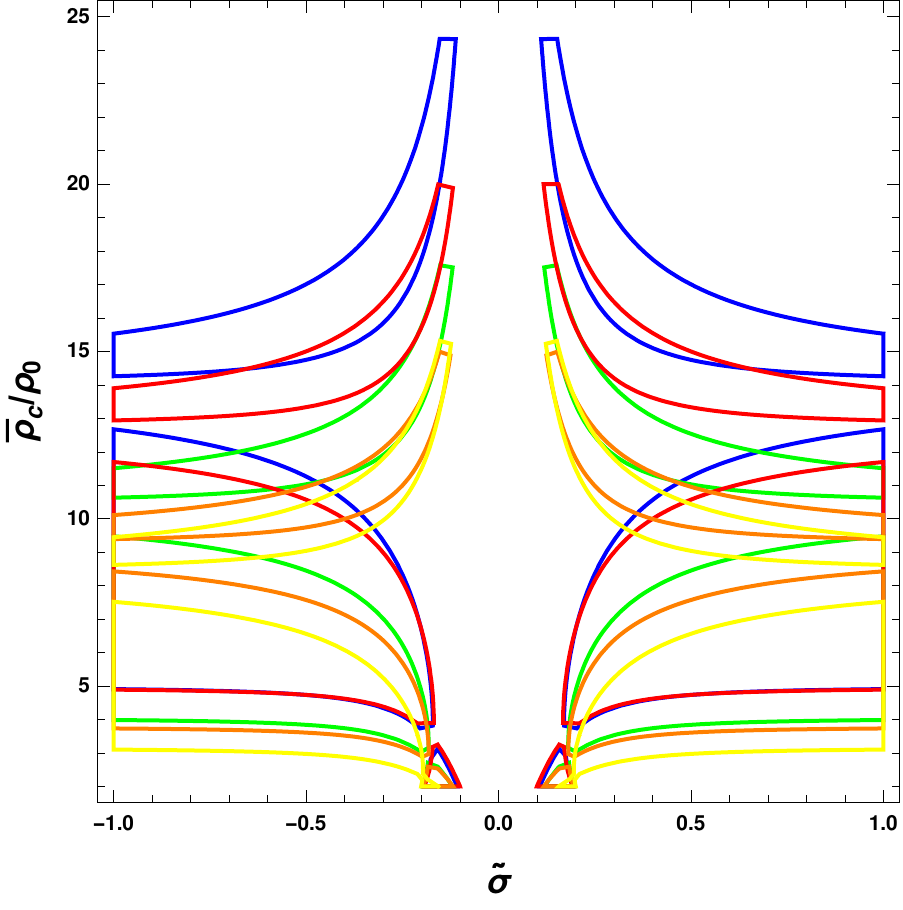} 
 \caption{\small The values of $\tilde{\cC}$ and $\overline{\rho}_c$ in which the condition $\overline{P}''(r=0)<0$ is satisfied for different EoS: FPS (with blue curve), SLy (with green curve), BSk19 (with red curve), BSk20 (with orange curve) and BSk21 (with yallow curve). The points, with the coordinate  $(\tilde{\cC},\overline{\rho}_c)$, inside the curves are the ones which can not use as a parameters of the model to obtain a physical NS. The curves are symmetric with respect to the axes $\overline{\rho}_c/\rho_0=0$ because  $\overline{P}''$ is a function of $\tilde{\cC}^2$.}
\label{Figmrho}
\end{figure}

For odd and even modes, the tidal love numbers $k_2^{E}$ and $k_2^{B}$ are proportional to the ratio $c_1/c_2$ and they are given by \cite{Damour:2009vw}
\begin{eqnarray}
k_2^E&=&\frac{8 }{5} \text{C}^5 (1-2 \text{C})^2 (2 \text{C} (y-1)-y+2)/\left(4 \text{C}^3 \left(2 \text{C}^2 (y+1)+\text{C} (3 y-2)-11 y+13\right)\right.\nonumber\\
&&\left.+2 \text{C} (3 \text{C} (5 y-8)-3 y+6)+3 (1-2 \text{C})^2 (2 \text{C} (y-1)-y+2) \log (1-2 \text{C})\right)\\
k_2^B&=&\frac{8 \text{C}^5 (2 \text{C} (2-y)+y-3)}{5 \left(2 \text{C} \left(2 \text{C}^3 (y+1)+2 \text{C}^2 y+3 \text{C} (y-1)-3 y+9\right)+3 (2 \text{C} (y-2)-y+3) \log (1-2 \text{C})\right)}
\end{eqnarray}
where $\text{C}= M/r_s$ is the compactness of the star and $y$  is evaluated at the surface of the star as $y=r_sH(r_s)/H'(r_s)$ and $y=r_sh(r_s)/h'(r_s)$ for the electric and magnetic tidal love number, respectively, by matching the interior and the exterior solutions. The interior solutions of the perturbed and non perturbed metrics as well as the scalar field are solved numerically by taking into account the following initial conditions
\begin{eqnarray}
&h=1+\cK\frac{ 16 \cK  \overline{\rho}_c-\cC^2 \left(\overline{\rho}_c-3 \overline{P}_c\right)^2}{6 \cR \left(4 \cK +\cC ^2 \left(\overline{\rho}_c-3 \overline{P}_c\right)\right)^2}r^2+ O\left(r^3\right),\label{series1}\\
&\frac{f}{f_0}= 1+\cK\frac{\overline{\rho}_c \left(8 \cK-5 \cC^2 \overline{\rho}_c\right)+6 \overline{P}_c \left(4 \cK +5 \cC^2 \overline{\rho}_c\right)-45 \cC^2 \overline{P}_c^2}{6 \cR  \left(4 \cK +\cC ^2 \left(\overline{\rho}_c-3 \overline{P}_c\right)\right)^2}r^2+ O\left(r^3\right),\label{series2}\\
&\overline{P}=\overline{P}_c+\frac{\cK  c_{m,0}^2 \left(\overline{P}_c+\overline{\rho}_c\right) \left(\overline{\rho}_c \left(5 \cC^2 \overline{\rho}_c-8 \cK \right)-6 \overline{P}_c \left(4 \cK +5 \cC^2 \overline{\rho}_c\right)+45 \cC^2 \overline{P}_c^2\right)}{3 \cR  \left(4 \cK  +\cC^2 \left(\overline{\rho}_c-3 \overline{P}_c\right)\right) \left(8 \cK  c_{m,0}^2-\cC^2 \left((3 c_{m,0}^2+1) \overline{P}_c+(1-5 c_{m,0}^2) \overline{\rho}_c\right)\right)}r^2+ O\left(r^3\right),\label{series3}\\
&h_0=h_{3}r^3,\qquad H=H_{2}r^2+ O\left(r^3\right),\label{series4}\\
&\delta\varphi  =\frac{\cC H_{2}\left((1-c_{m,0}^2) \overline{\rho }_c \left(8 \cK -\cC ^2 \overline{\rho }_c\right)+3 \cC ^2 (1-9 c_{m,0}^2) \overline{P}_c^2+2 (3 c_{m,0}^2+1) \overline{P}_c \left(4 \cK +\cC^2 \overline{\rho }_c\right)\right)}{6 \left(4 \cK +\cC ^2 (\overline{\rho }_c-3 \overline{P}_c)\right) \left(4 \cK c_{m,0}^2-\cC^2 c_{m,0}^2 (9 \overline{P}_c+\overline{\rho }_c)+\cC^2 (\overline{P}_c+\overline{\rho }_c)\right)}r^3+ O\left(r^3\right),\label{series5}
\end{eqnarray} 
with $c_{m,0}^2=c_{m}^2(r=0)$, to avoid numerical instability. In order to get a well behave neutron star, we impose $\overline{P}''(r=0)<0$, where the values of $\tilde{\cC}=(\cR/(\cK\rho_0^2))^{(1/2)}\cC$ and $\overline{\rho}_c$ that satisfied this condition are resumed in Fig.\ref{Figmrho}. Note that $\rho_0$ is the nuclear density and it is defined in the next section \ref{Sec4}.

\section{Numerical analysis}\label{Sec4}
\begin{figure}[h]
\centering 
\includegraphics[scale=0.8]{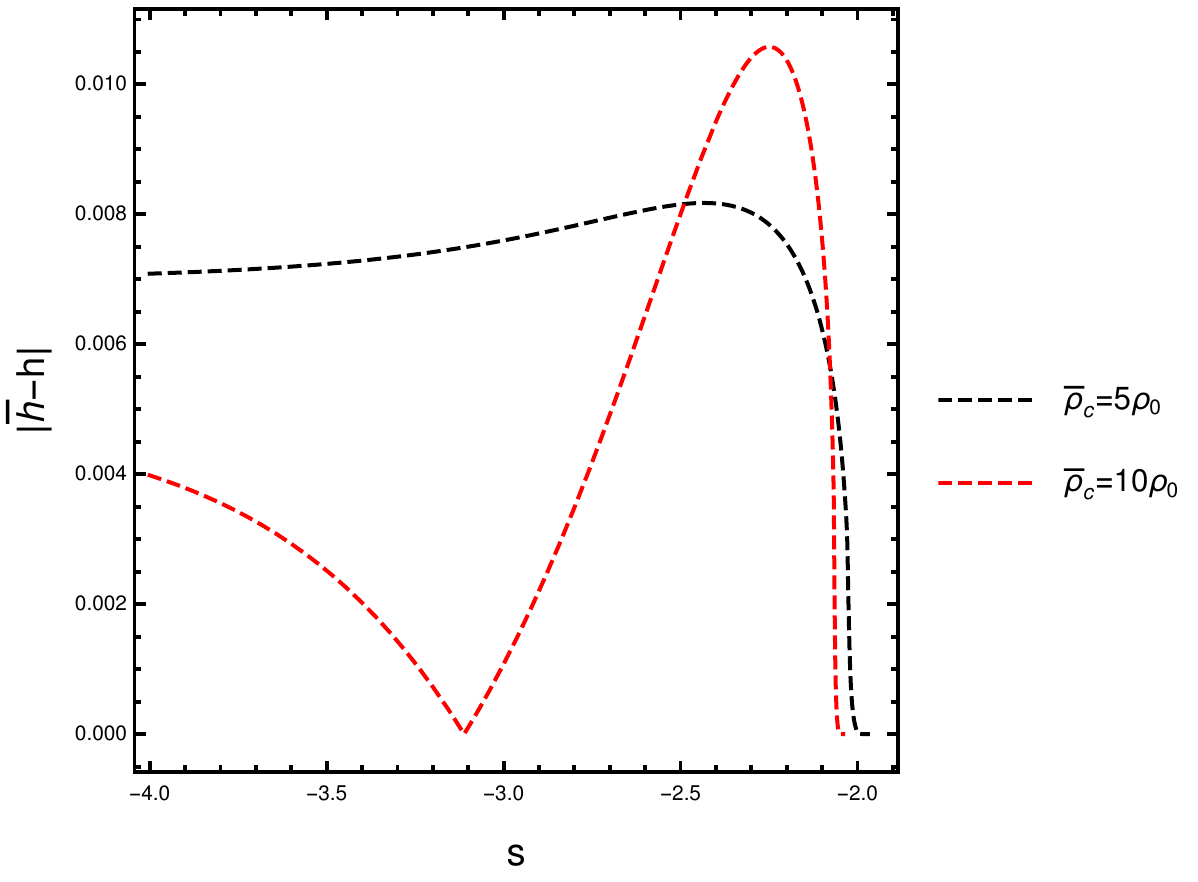} 
 \caption{\small The difference between the background metrics in the Jordan  and Einstein frames as function of $s$ for $\Tilde{\cC}=0.03$ and by using the EoS BSk21.}
\label{FighP}
\end{figure}
\begin{figure}[h]
\centering
\includegraphics[scale=0.82]{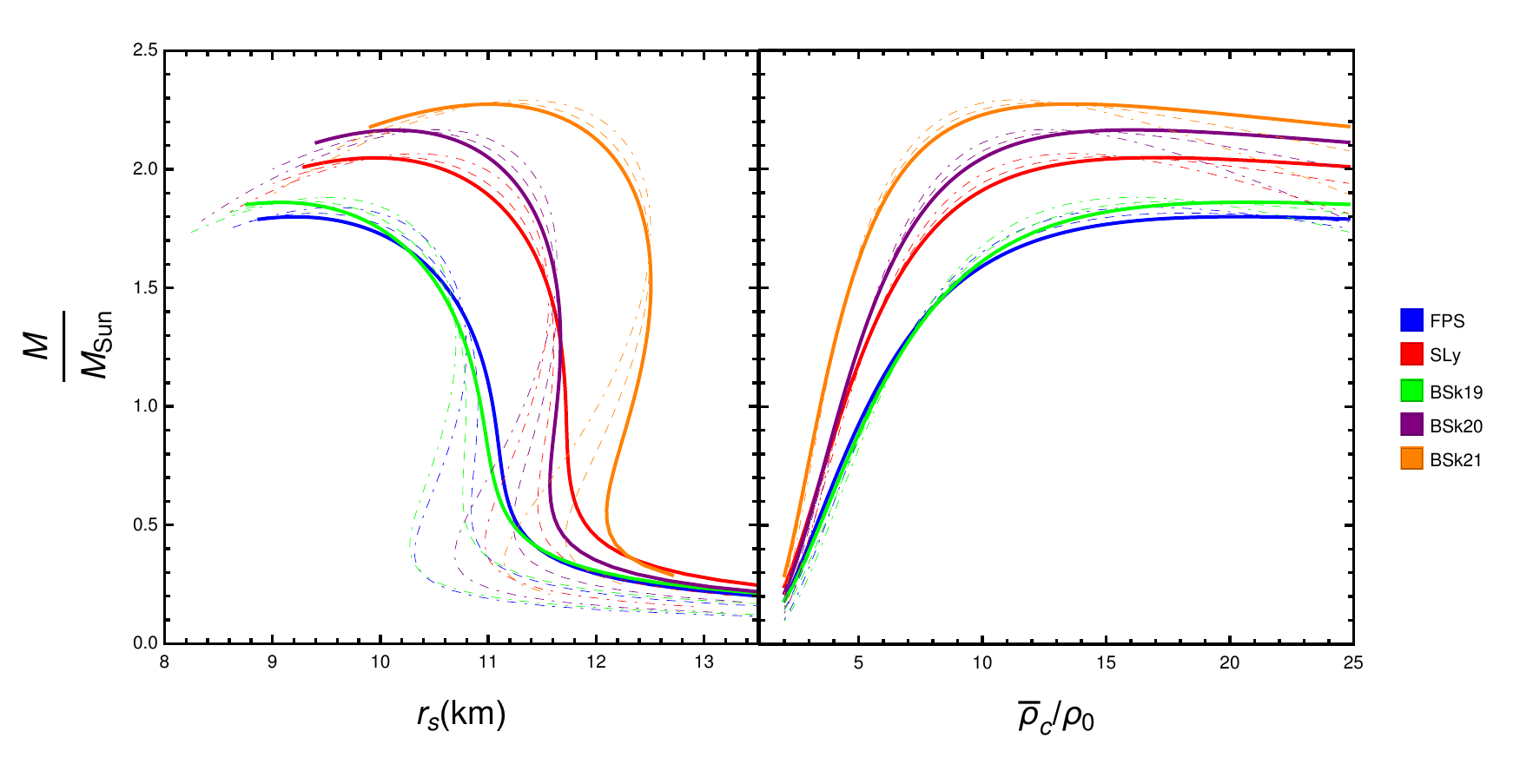} 
\caption{\small The mass-radius (left graph) and mass-central density (right graph)  relations of NS   using the EoSs: FPS, SLy, BSk19, BSk20 and BSk21 for: $\Tilde{\cC}=0$ (Solide lines),  $\Tilde{\cC}=0.03$ (Dashed  lines),  $\Tilde{\cC}=0.045$ (DotDashed  lines). The colours and curves used here are the same in Ref.\cite{Boumaza:2021fns}. $M_{Sun}$ corresponds to the mass of NS.}
\label{MRs}
\end{figure}

{\hskip 2em}To solve the   background and perturbation equations for the particular cases (\ref{functions}), we follow the same steps described in Refs.\cite{Boumaza:2021fns,kase2019neutron}. In these papers, the authors integrate numerically the equations corresponding to their models by  defining the dimensionless quantities:
\begin{eqnarray}
\tilde{\rho}=\frac{\overline{\rho}}{\rho_0},\quad \tilde{P}=\frac{\overline{P}}{\rho_0 c^2}\quad \text{and}\quad s=Ln\frac{r}{r_0},\label{var1}
\end{eqnarray}
where 
\begin{eqnarray}
\rho_0 = m_n n_0=1.6749\times 10^{14}\; g.cm^{-3},\qquad r_0=\frac{c}{\sqrt{G \rho_0}}=89.664 \; km.
\end{eqnarray}
And $m_n=1.6749\times 10^{-24}g$ and $n_0 = 0.1 fm^{-1}$ are, respectively, the neutron mass and the typical number density of NSs. The constant $\cK$ can disappear from the equations  after rescaling the scalar field and the constant $\cC$ as:
\begin{eqnarray}
\delta\varphi\rightarrow\sqrt{\frac{\cR}{\cK}}\delta\varphi, \qquad\varphi\rightarrow\sqrt{\frac{\cR}{\cK}}\varphi\qquad \text{and}\qquad\cC\rightarrow \sqrt{\frac{\cK }{\cR\rho_0^2}} \tilde{\cC}.\label{var2}
\end{eqnarray}
In order to write  Eq. (\ref{eH})  and (\ref{eHphi}) as a system of differential equations of the first order we use the variables:
\begin{eqnarray}
y_{odd}=\frac{rh_0'}{h_0},\quad y_{even}=\frac{rH'}{H},\quad y_{\delta\varphi'}=\frac{\delta\varphi'}{H}\quad\text{and}\quad y_{\delta\varphi}=\frac{\delta\varphi}{H}.\label{y}
\end{eqnarray}
{\vskip 2em }Thus, the model contains only two parameters which are the central energy density $\overline{\rho}_c$ and the constant $\Tilde{\cC}$. We integrate numerically the equations (\ref{eh}),  (\ref{ef}), (\ref{eP}), (\ref{eH})  and (\ref{eHphi}) from the center to the surface of the star for the special case (\ref{spetialfuctions}) and using the EoSs: FPS, SLy, BSk19, BSk20 and BSk21.  At the center of the star, we consider the  development derived in Eqs. (\ref{series1}-\ref{series5}) as initial conditions of the energy density,  perturbed and non-perturbed metrics. In addition, the radius of the star $r_s$ is determined numerically by the equation $\overline{P}(r_s)=0$, but in our resolution we take the approximate value of $\overline{P}(r_s)=10^{-10}\rho_0$ to avoid numerical instabilities. As we have shown, in our model one can solve the equations of motion analytically at the background and perturbed level. The constants of integration that appear in these solution are determined by matching the exterior and interior solutions, i.e. we impose $f_{in}(r_s)=f_{out}(r_s)$, $h_{in}(r_s)=h_{out}(r_s)$, $y_{even,in}(r_s)=y_{even,out}(r_s)$, $y_{odd,in}(r_s)=y_{odd,out}(r_s)$ and $y_{\delta\varphi,in}(r_s)=y_{\delta\varphi,out}(r_s)$, thus one can calculate the mass and tidal love numbers (even and odd type) of the NS.  
\begin{figure}[h]
\centering 
\includegraphics[scale=0.7]{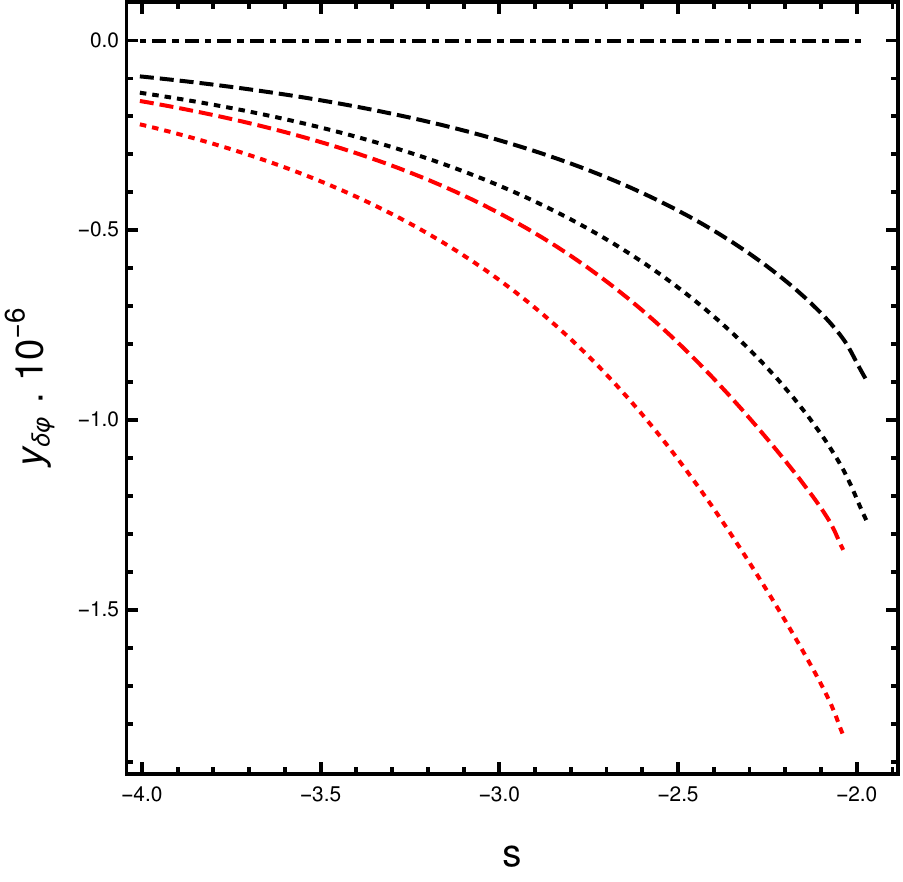} \includegraphics[scale=0.7]{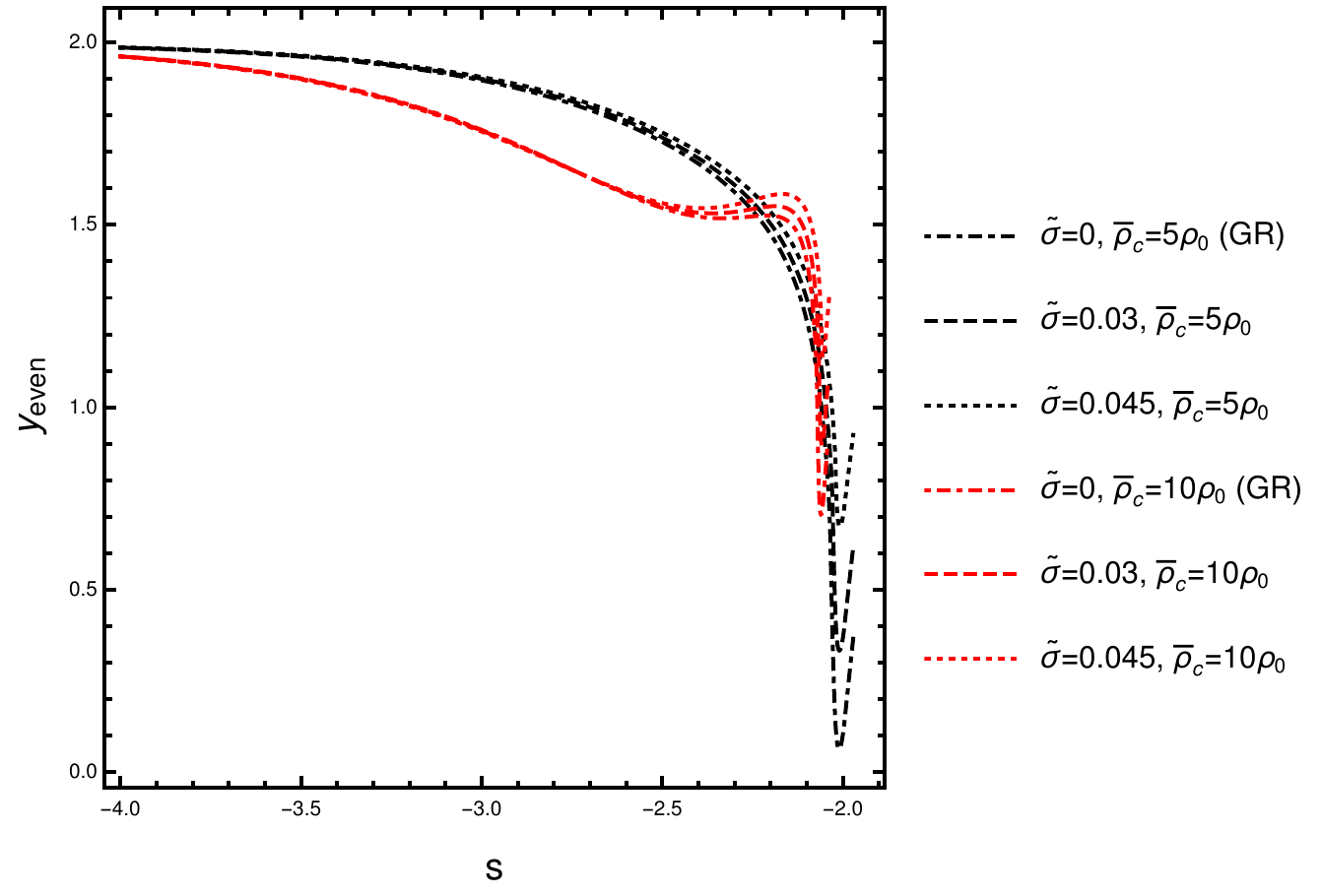}
 \caption{\small The variation of the quantities $y_{even}$ and $y_{\delta\varphi'}$  as function of $s$ for, using the EoS BSk21, for the parameters ($\Tilde{\cC},\overline{\rho}_c$): ($0,5\rho_0$) (with black dotdashed curve), ($0.03,5\rho_0$) (with black dashed curve), ($0.045,5\rho_0$) (with black dotted curve), ($0,10\rho_0$) (with red dotdashed curve),($0.03,10\rho_0$) (with red dashed curve) and ($0.045,10\rho_0$)(with red dotted curve).}
\label{Figheven}
\end{figure}
{\vskip 2em } The numerical solutions will allow us to compare between the metrics, scalar field and pressure in the Jordan and Einstein frames, as an example: In Fig.\ref{FighP}, we show the variation of the absolute value of $h-\overline{h}$ with respect to $s$  using BSk21 and the parameters $\Tilde{\cC}=0.03$ for two central densities ($\overline{\rho}_c=\{5\rho_0,\;10\rho_0\}$). As expected, the two metrics are not equal at the center of the star but coincide at the surface  because $\varphi'(r=r_s)=0$. In Fig.\ref{Figheven} and Fig.\ref{Figodd}, we plot the radial evolution of the dimensionless quantities of Eq.(\ref{y}) using the parameters and the EoS mentioned in the figures, where we observe that these functions are quasi-identical when $r\rightarrow 0$ and the deviation from GR start to appear when $r$ is close to $r_s$. We observe also the deviation is significant  and the behaviour  of $y_{even}$ changes when we vary $\overline{\rho}_c$. Because of these deviations, one can expect that the mass, radius as well as the tidal love number of the two types will be different from those we found in GR. At the end of the paragraph, we mention that the values of $\Tilde{\cC}$ are chosen to satisfy the constraints $P''(0)>0$ which are presented in Fig.\ref{Figmrho}.
 \begin{figure}[h]
\centering 
\includegraphics[scale=0.7]{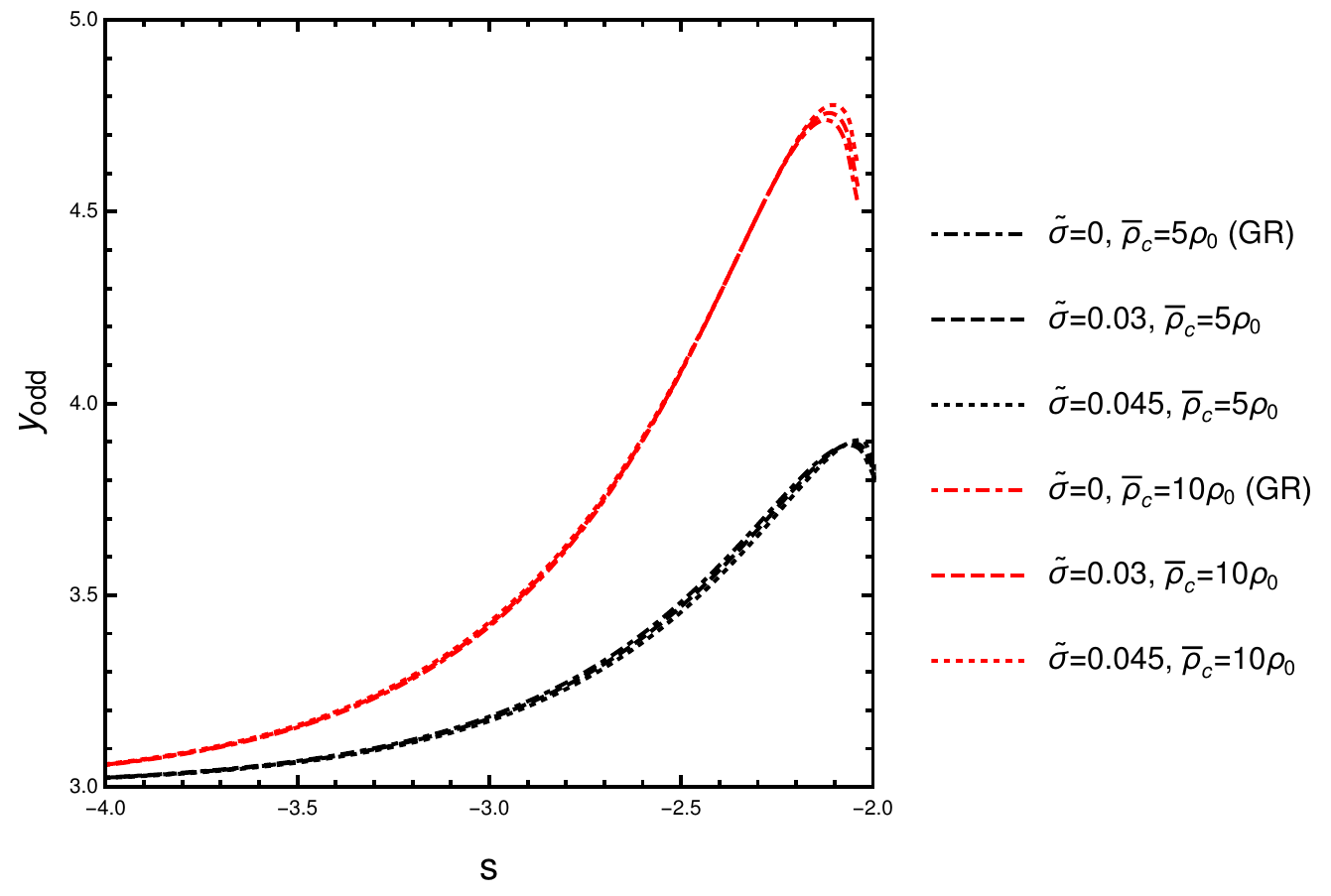}\includegraphics[scale=0.7]{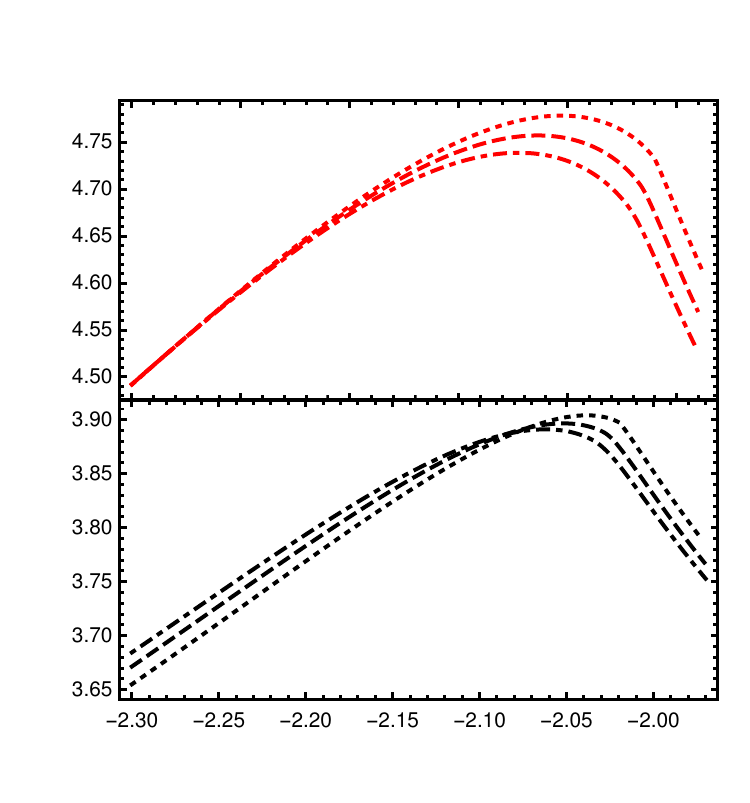} 
 \caption{\small In the left graph, we plot the variation of  $y_{odd}$ as function of $r$ for, using the same EoS, the parameters and notations in Fig.\ref{Figheven}. In the right graph, we plot a zoom of the graphs to show the differences between the curves at the surface of the star.}
\label{Figodd}
\end{figure}
{\vskip 2em }In right plane of Fig.\ref{MRs}, we compare our model with GR, which can be defined by $\Tilde{\cC}=0$,  where we have plotted Mass-radius relation for different realistic equations of state and for the value $\Tilde{\cC}=\{0,0.03,0.045\}$, by varying $\overline{\rho}_c$ from $2\rho_0$ to $25\rho_0$. The plots show that the maximum mass and its  radius increase when the value of $\Tilde{\cC}$ increases for the five equations of state. The plot also showed that the mass of the NS in our model, for different values of $\Tilde{\cC}=0$,  can be lower and bigger than that in GR for low and high central densities.  We observe also that the maximum of the mass can exceed the value $2M_{Sun}$, where $M_{Sun}$ is the mass of the sun, for the SLy,  BSk20 and BSk21 EoSs but it is not the case for FPS and BSK19 which can be ruled out  by the observation the pulsar PSR J1614-2230 \cite{Demorest:2010bx}. However, one can revive these EoS by adding a torsion term coupled to scalar field, as it has been done in Ref.\cite{Boumaza:2021fns}. Moreover, if we do so, the NS mass can  exceed the value $2.5 M_{Sun}$, which is the mass of the compact object calculated from the GW190814  event. In left plane of Fig.\ref{MRs}, using the same values of $\Tilde{\cC}$ and the same EoSs,  we plot  the variation of NS mass with respect the central density. We can see  that the deviation from GR becomes important at high central density and the maximum mass is in the range $\overline{\rho}_c\in[5,10]$. 

\begin{figure}[h]
\centering
 \includegraphics[scale=0.7]{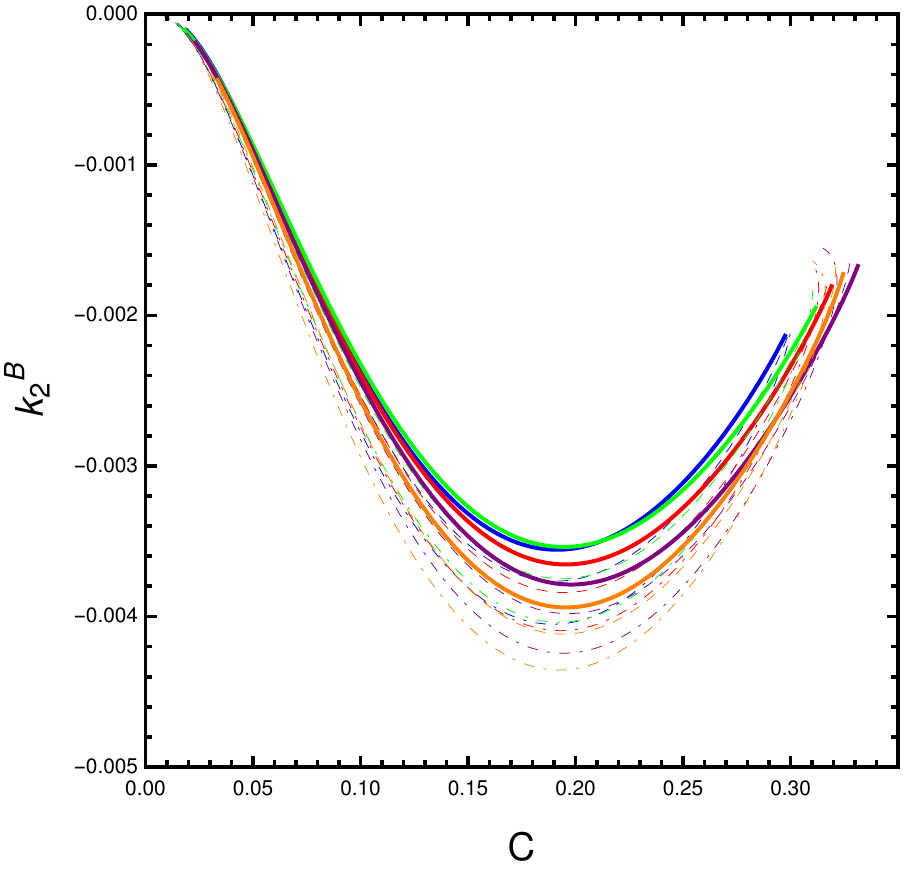}\includegraphics[scale=0.7]{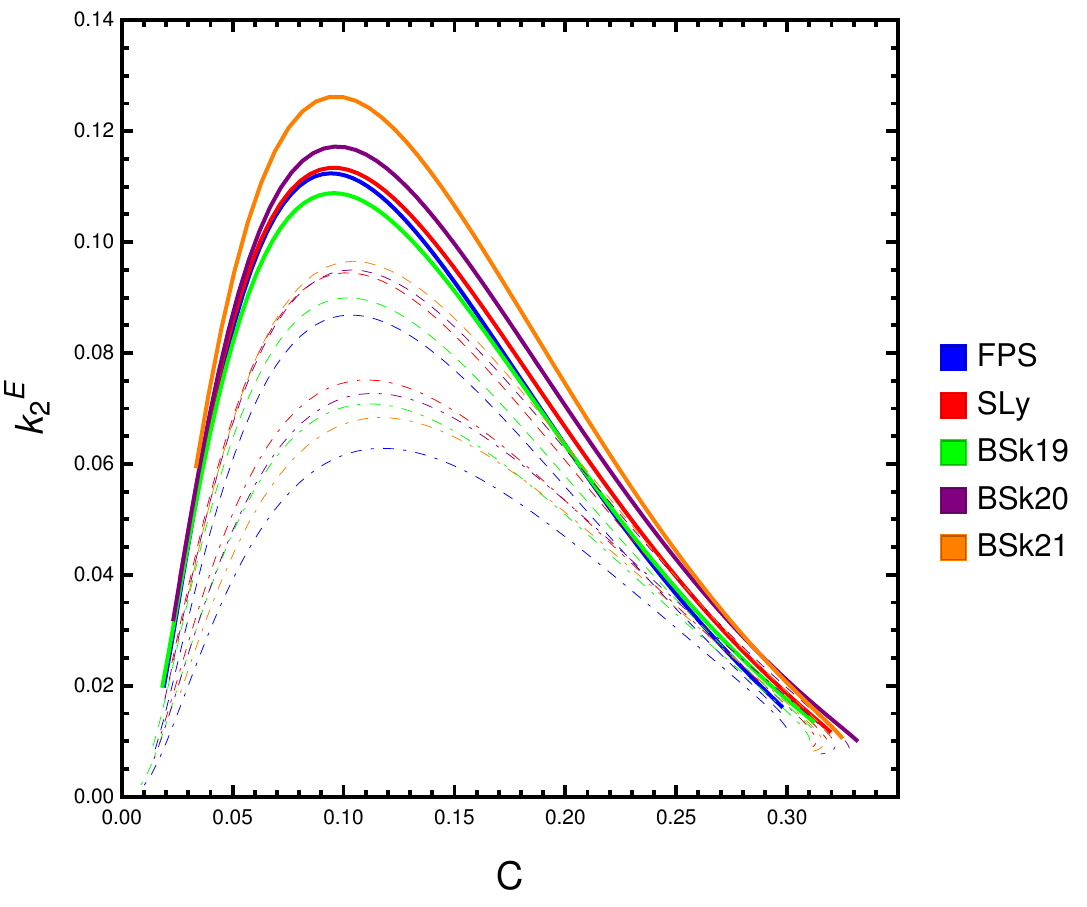} 
\caption{\small The variation of magnetic (right graph) and electric (left graph) tidal love number versus the compactness of NS for the cases in Fig.\ref{MRs}.}
\label{kC}
\end{figure}

{\vskip 2em }We present, in the graphs of  Fig.\ref{kC}, the variation of magnetic  and electric  tidal love number with respect to $C$ for $\Tilde{\cC}=\{0,0.03,0.045\}$ and for FPS, SLy, BSk19, BSk20 and BSk21 EoSs. We notice that the polar TLNs can differ from GR approximately 3 times by varying the parameter $\Tilde{\cC}$ and  the EoS while the axial love number can vary up to roughly $15\%$ for the range of $\Tilde{\cC}$ considered in the plot. In fact, this is due to the small deviations of $y_{odd}$  and the considerable deviations of the $y_{even}$  at the surface which  can be seen in Fig.\ref{Figheven} and Fig.\ref{Figodd}. The axial contribution to the gravitational waves is neglected compared to the polar contribution and it is also the case for higher value of $l$ as long as  the  change  of  the  phase  in  the  gravitational wave signal is concerned \cite{Damour:2009vw}. Thus, constraining the parameter $\Tilde{\cC}$ through the axial TLNs  would be difficult to observe in practice.

\begin{figure}[h]
\centering
 \includegraphics[scale=0.7]{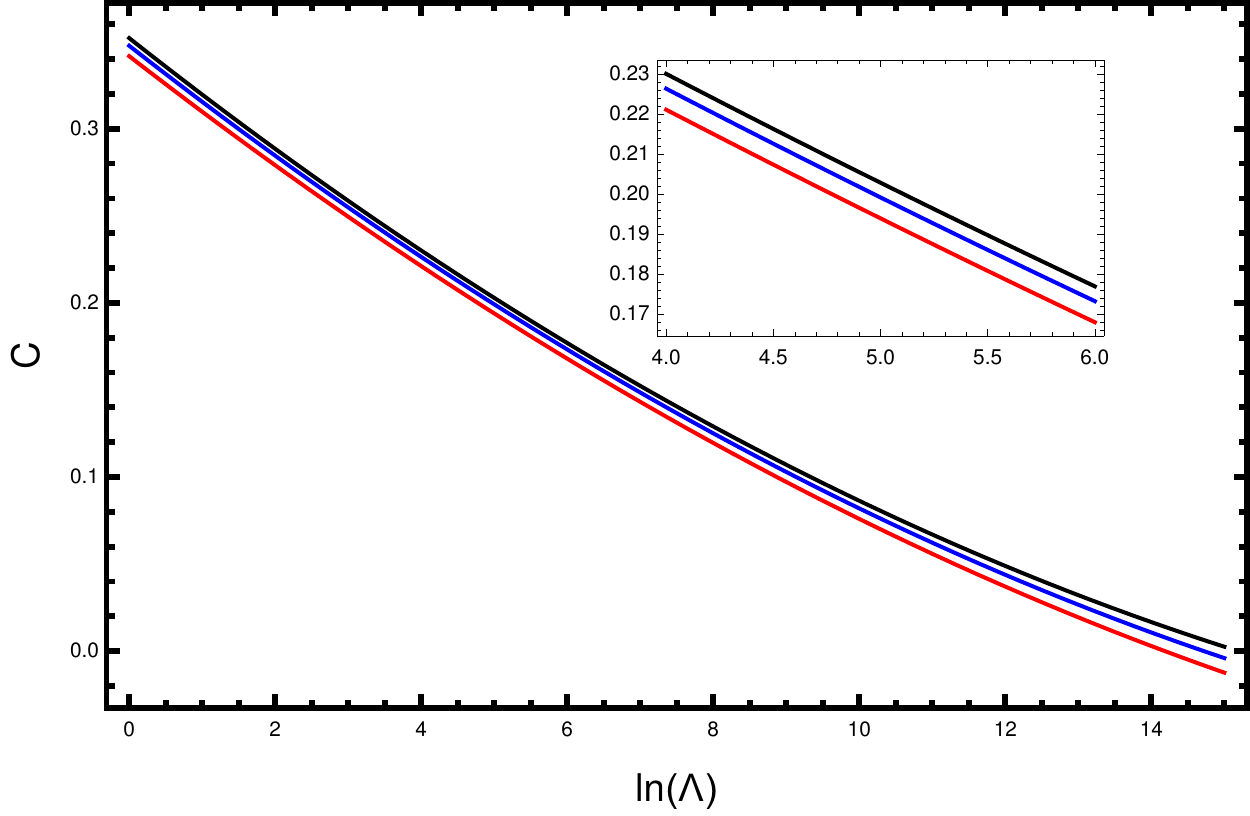}
\caption{\small In this figure, we present the universal relation $C$-$\Lambda$ using the coefficients of the table \ref{table2} for BSk21 EoS with $\Tilde{\cC}=\{0\;(\text{Black line}),0.03\;(\text{Blue Line}),0.045\;(\text{Red Line})\}$. }
\label{Clambda}
\end{figure}

{\vskip 2em }In order to compare  our models with the G170817 event, which constrains  the value of the tidal deformability of NS $\Lambda=k_2^E C^{-5}(2/3)$ to $190^{+390}_{-120}$ ($90\%$ CL) for a $1.4 M_{sun}$, we adopt the universal relation between $\Lambda-C$ where it appears for the first time in Ref.\cite{Maselli:2013mva} as
\begin{eqnarray}
C=b_0+b_1 \ln\Lambda+ b_2 \ln\Lambda,\label{CLambda}
\end{eqnarray}
where $b_0$, $b_1$ and $b_2$ are constants determined by fitting this relation with our model for different EoSs  values of $\Tilde{\cC}$. From (\ref{CLambda}), it is easy to calculate the radius of the NS using its tidal deformability with $90\%$ CL. The coefficients in (\ref{CLambda}) and the corresponding radius are illustrated in table \ref{table2} and table \ref{table3}. For example, according to the universal  relation (\ref{CLambda}), the radius of the neutron star, for FPS EoS with same masses as GW170817 and $\Tilde{\cC}=0.03$, is $11.16^{+2.07}_{-1.47}\,km$ ($90\%$ CL), i.e.  the maximum radius of this star is $13.23\,km$ and the minimum is $9.69\,km$. Therefore, this particular model satisfies the GW170817 observation because the minimum of the radius provided from the mass-radius relation in the range $M\in [M_{Sun},2M_{Sun}]$ is $10.87\,km$. In addition, in GR the BSk21 EoS is ruled out since the minimum radius for such star in $M\in [M_{Sun},2M_{Sun}]$ is $r_{min}=12.44$ which is not included in the interval $[12.31,9.25]\,km$. However, the realistic equation of state BSk21 remains viable when $\Tilde{\cC}=\{0.03,0.045\}$, following the GW170817 observation. Finally, we show in Fig.\ref{Clambda}  the fitted $C$-$\ln \Lambda$ relation only for the BSk21 EoS, to have a better visibility but with different values of $\Tilde{\cC}$, where we can see the small effect of the parameter $\Tilde{\cC}$ on the universal relation.

\begin{table}[h]
\centering
\scalebox{0.7}{
\begin{tabular}{||c || c c c ||c c c ||c c c|| } 
\hline\hline
Model                         & & GR &  & & $\Tilde{\cC}=0.03$ & & & $\Tilde{\cC}=0.045$& \\
\hline\hline
Coefficients                  & $b_0$ & $b_1$ & $b_2.10^{-4}$ & $b_0$ &  $b_1$ & $b_2.10^{-4}$& $b_0$ & $b_1$ &   $b_2.10^{-4}$\\
\hline\hline      
FPS                   & 0.355786 & -0.0368011 & 9.53068 & 0.354140& -0.0373946 & 9.86851 &  0.351830 & -0.0382787 & 10.4072\\
SLy                   & 0.354069 & -0.0354115 & 8.43473 & 0.350761& -0.0351550 & 8.27603 &  0.346387 & -0.0349036 & 8.17221\\
BSk19                 & 0.35666  & -0.0369933 & 9.62134 & 0.354106& -0.0370604 & 9.68461 &  0.350693 & -0.0372198 & 9.84603\\
BSk20                 & 0.355186 & -0.0350321 & 8.10727 & 0.350749& -0.0346123 & 7.87723 &  0.344713 & -0.0341463 & 7.69452\\
BSk21                 & 0.352046 & -0.0331128 & 6.54154 & 0.347652& -0.0328264 & 6.25585 &  0.341676 & -0.0325246 & 5.94878\\
\hline\hline 
\end{tabular}}
\caption{\small The best fitting coefficients  determined from eq. (\ref{CLambda})  for GR and our model.}
\label{table2}
\end{table}

\begin{table}[h]
\centering
\scalebox{0.7}{
\begin{tabular}{||c || c c  ||c c ||c c || } 
\hline\hline
Model                         & & GR & &  $\Tilde{\cC}=0.03$ &  & $\Tilde{\cC}=0.045$ \\
\hline\hline
Coefficients                 &  $r_{min}$& $r_s$(km) &$r_{min}$ & $r_s$(km)& $r_{min}$& $r_s$(km)\\
\hline\hline      
FPS                   &$10.92$ &$10.93^{+1.96}_{-1.40}$&$10.87$ &$11.16^{+2.07}_{-1.47}$&$10.80$  &$11.50^{+1.94}_{-1.68}$\\
SLy                   &$11.67$ &$10.79^{+1.90}_{-1.35}$&$11.61$ &$10.92^{+1.94}_{-1.38}$&$11.51$  &$11.12^{+2.00}_{-1.42}$\\
BSk19                 &$10.80$ &$10.93^{+1.97}_{-1.40}$&$10.76$ &$11.09^{+2.03}_{-1.44}$&$10.70$  &$11.32^{+2.12}_{-1.50}$\\
BSk20                 &$11.17$ &$10.66^{+1.85}_{-1.32}$&$11.59$ &$10.82^{+1.90}_{-1.35}$&$11.49$  &$11.06^{+1.96}_{-1.39}$\\
BSk21                 &$12.44$ &$10.52^{+1.79}_{-1.27}$&$12.35$ &$10.72^{+1.86}_{-1.31}$&$12.27$  &$11.02^{+1.98}_{-1.38}$\\
\hline\hline 
\end{tabular}}
\caption{\small The table shows the radius of NS calculated from eq. (\ref{CLambda})  and the minimum radius determined from mass-radius relation  for GR and our model.}
\label{table3}
\end{table}

\section{Conclusion}\label{Sec5}
{\hskip 2em}In this study, we have investigated the physical proprieties of the neutron star where we have considered two metrics linked by a disformal transformation for five realistic EoSs. We have chosen a model in which the shift symmetry and the constraint from the GW170817 event is verified. These assumptions have reduced our model to the case (\ref{functions}) which corresponds to the conformal transformation. The equations of motion that describe a static NS were calculated at the background and at the perturbed level as well as the asymptotic behaviours of the metric, pressure and scalar field have been examined when $r\rightarrow 0$ and $r>r_s$. The particular functions (\ref{spetialfuctions}) give a vanishing scalar field at the exterior of the star and hence GR solutions are found and the two frames become identical.\\

A numerical analysis of the equations has been performed, in order to examine the radial behaviour of the metrics, for different values of $\Tilde{\cC}$ and for different equations of state. From the numerical  solutions, we observe the deviation of our model from GR, which the case $\Tilde{\cC}= 0$, when $\Tilde{\cC}\neq 0$. Because this deviations, the mass, the radius and even the TLN (even and odd modes) of the star have different values from those in GR. Additionally, we found that the TLN for even modes are much smaller than GR while axial TLNs are slightly different in contrast with the results we found in Ref.\cite{Yazadjiev:2018xxk}. This has motivated us to compare our results with the observations from the GW170817 event using the universal relation (\ref{CLambda}). Finally, we found that our model is compatible with the observations for $\Tilde{\cC}=\{0.03,0.045\}$ and for the five realistic EoSs while in GR only BSk21 is ruled out.

\appendix
\section{Coefficients in the action $S^{(2)}_{vac}$}\label{A1}

{\hskip 2em}The explicit expressions of the background-dependent coefficients $p_i$ are given by
\begin{eqnarray}
&p_{0}=-\frac{1}{8} f h r^2\FK -\frac{f \cR  }{8 h^2}\left(2 r h'+h^3-h\right),\quad p_{1}=\frac{f \cR r^2}{4 h},\quad p_{2}=-2p_{1},\quad p_{3}=\frac{f \cR  r }{4 h^2}\left(3 h-r h'\right),\nonumber\\
&p_{4}=\frac{f h \cR  l(l+1)}{2},\quad p_{5}= -\frac{f r^2 \FK_X \varphi '}{h},\quad p_{6}= -\frac{1}{4} f \FK h r^2-\frac{f \cR }{8 h^2}\left(2 h^3 l(l+1)+4 r h'+h^3-2 h\right),\nonumber\\
&p_{7}=-\frac{f r^2 h}{4 }\left(\FK -2 \FK_X X\right)-\frac{f \cR }{4 h^2}\left(h^3 (l(l+1)+1)-2 r h'+h\right),\quad p_{8}=\frac{1}{4} f h \cR  (l (l+1)+1),\nonumber\\
&p_{9}=\frac{f \cR  l (l+1) }{2 r}\left(r h'+h\right),\quad p_{10}=\frac{\cR }{8 h}\left(6 r f'-f h^2+3 f\right)+\frac{f r^2 h }{8 }\left(4  \FK_X X+4 \FK_{XX} X^2-\FK \right),\nonumber\\
&p_{11}=-\frac{\cR  r }{4 h}\left(r f'+f\right),\quad p_{12}=-\frac{f r^2  \varphi'}{h}\left(\FK_X+2 \FK_{XX} X\right),\quad p_{13}=\frac{fh r^2 }{4}\left(\FK -2 \FK_X X\right)\nonumber\\
&+\frac{\cR}{8 h}\left(f \left(h^2 (2 l (l+1)+1)-2\right)-4 r f'\right),\quad p_{14}=-\frac{f h\cR (l (l+1)+1)}{4},\quad p_{15}= -\frac{h \cR l (l+1) }{2 r}\left(r f'+f\right),\nonumber\\
&p_{16}=l (l+1)\left(\frac{h \cR  }{2 r^2}\left(2 r f'+f\right)+f h^3 \left(\FK_X X-\frac{1}{2} \FK \right)\right),\quad p_{17}=\frac{f h \cR  (2 l (l+1)+1)}{4},  \nonumber\\
& p_{18}=-\frac{f h \cR  (l (l+1)+1)}{2},\quad p_{19}=-2 f h l (l+1) \FK_X \varphi',\quad p_{20}=2p_{13},\nonumber\\
&p_{21}=-\frac{ \cR  r }{8 h^2}\left(f' \left(h-r h'\right)+h r f''-f h'\right)-\frac{f \FK h r^2}{8},\quad p_{22}=\frac{f r^2\FK_X \varphi '}{h},\quad p_{23}=f h l(l+1) \FK_X,\nonumber\\
&\quad p_{24}=p_{1}, \quad  p_{25}=-p_{12}.
\end{eqnarray} 

\section{Perturbation equation of matter conservation}\label{A2}
{\vskip 2em} Let's define the first order perturbation of  the energy momentum tensor, in the Jordan frame, by 
\begin{eqnarray}
\delta\overline{T}^{\mu\nu(m)}=\frac{2}{\sqrt{-\overline{g}^{(0)}}}\frac{\delta\left(\sqrt{-\overline{g}}\Lm^{(2)}\right)}{\delta\overline{g}_{\mu\nu}^{(1)}}.
\end{eqnarray}
In the case $H_5=0$, the nonvanishing components for even modes are calculated by
\begin{eqnarray}
\delta\overline{T}^{tt}&=& \frac{\overline{\rho}  \C_X \varphi' \left(H_2 \varphi '-2 \delta \varphi '\right)+\C h^2( \delta \overline{\rho}  + H_0 \overline{\rho} )}{\C^2 f^2 h^2}Y_{l0}, \\
\delta\overline{T}^{rt}&=& \frac{H_1 \overline{P}}{\C f^2 h^2} Y_{l0}, \\
\delta \overline{T}^{rr}&=& \frac{H_2 h^2\left(\overline{P} \C_X X-\C \overline{P}\right)-2 \overline{P} \C_X \delta \varphi' \varphi '+\C \delta \overline{P} h^2}{\C^2 h^4}Y_{l0},  \\
\delta\overline{T}^{\theta\theta}&=& \frac{\overline{P} \C_X \varphi ' \left(H_2 \varphi '-2 \delta \varphi '\right)+\C h^2( \delta \overline{P} -  \overline{P} H_3)}{\C^2 h^2 r^2}Y_{l0}, \\
\delta \overline{T}^{\phi\phi}&=&  \frac{\overline{P} \C_X \varphi ' \left(H_2 \varphi '-2 \delta \varphi '\right)+\C h^2(\delta \overline{P} -\overline{P} H_4)}{\C^2 h^2 r^2\sin^2\theta } Y_{l0}.
\end{eqnarray}
while the nonvanishing components of the equation of matter conservation are 
\begin{eqnarray}
\frac{\delta \overline{P}'}{\overline{P}+\overline{\rho}}&=&\frac{1}{2}H_0'+\frac{\C_X X }{2 \C }H_2'-\frac{\C_X   \varphi '}{\C h^2}\delta \varphi '-\frac{\delta \overline{P}+\delta \overline{\rho}}{\overline{P}+ \overline{\rho}}  \left(\frac{\C_X  X'}{2\C}-\frac{f'}{f}\right)\nonumber\\
&&-\frac{\C  \C_X \left(h \varphi ''-2 h' \varphi '\right)-2X( \C \C_{XX} - \left(\C_X\right)^2)  \left(h' \varphi '-h \varphi ''\right)}{\C^2 h^3}\delta\varphi '\nonumber\\
&&+\frac{ X' \left(\C  \C_X+\C \C_{XX}  X-\left(\C_X\right)^2 X\right)}{2\C^2 }H_2,\label{eqm2}
\end{eqnarray}
and 
\begin{eqnarray}
\frac{\delta \overline{P}}{\overline{P}+\overline{\rho}}&=&\frac{1}{2}H_0-\frac{\C_X   \varphi '}{\C h^2}\delta \varphi ''+\frac{\C_X X }{2 \C }H_2,\label{eqm3}
\end{eqnarray}
where we have used  Eq. (\ref{eP}) to eliminate $\overline{P}'$. If we replace $H_0$ and $H_2$ by $H$ in Eq. (\ref{eqm3}) and we solve the resulting equation for $\delta\overline{P}$ then  we insert  the expression of $\delta\overline{P}$ in Eq. (\ref{eqm2}), the expressions of $\delta\overline{P}$ and $\delta\overline{\rho}$ are given by
\begin{eqnarray}
\frac{\delta\overline{P}}{\overline{P}+\overline{\rho}} &=& \frac{\C +\C_X X}{2\C}H-\frac{\C_X \varphi'}{\C h^2}\delta\varphi',\label{deltaP}\\
\delta\overline{\rho}&=&\frac{ \delta\overline{P}}{c_m^2}\label{deltarho}.
\end{eqnarray}
where $c_m^2=\partial\overline{P}/\partial\overline{\rho}$.

\section{$e_i$, $q_i$, $Y_i$ and $Z_i$ coefficients} \label{A3}
In this appendix, we show the expressions of the coefficients motioned in Sec.\ref{Sec2}.
\subsection*{$e_i$:}
\begin{eqnarray}
&e_0=2p_{11},\; e_1=-p_2,\; e_2=p_{12}+\C_X^2\frac{f r^2X\varphi '}{2 h} \left( (9\overline{P}+\overline{\rho} ) -\frac{(\overline{P}+\overline{\rho} ) \varphi '}{2 c_m^2 h}\right)-\C\C_X\frac{(3\overline{P}+\overline{\rho} )f r^2  \varphi '}{2 h},\nonumber\\
&e_3=p_7+2p_{10}-p_2'+f h r^2 X^2 \C_X^2 \left(\frac{\overline{P}+\overline{\rho} }{4 c_m^2}-\frac{1}{4} (9\overline{P}+\overline{\rho} )\right)+f h r^2 X \C_X \left(\frac{\C (\overline{P}+\overline{\rho} )}{4 c_m^2}+\frac{\C \overline{P}}{2}\right)\nonumber\\
&+\frac{1}{4} \C^2 f h r^2 (\overline{\rho}-\overline{P}),\; e_4= \C f hr^2 P  \C_X X+2 p_{13}+p_{14},\; e_5=2p_1,\; e_6=2p_3,\; e_7=p_2,\nonumber\\
&e_8=p_5-\C f r^2 \C_X\frac{ \varphi ' (2 c_m^2 \overline{\rho} +\overline{P}+\overline{\rho} )}{2 c_m^2 h},\; e_9=\frac{\C f r^2 (2 c_m^2 \overline{\rho} +\overline{P}+\overline{\rho} ) \left(\C_X X+\C \right)}{4 c_m^2 }+2 p_0+p_7\nonumber\\
&e_{10}=2 p_6+p_8.
\end{eqnarray}
\subsection*{$q_i$:}
\begin{eqnarray}
&q_0=p_1,\; q_1=-p_24',\; q_2=2 p_1'-p_3-p_{11},\;q_3=p_{22}-\frac{f r^2 \varphi '}{2 h}\C\C_X (3\overline{P}+\overline{\rho} ),\nonumber\\
&q_4=\frac{1}{4} \C^2 f h r^2 (\overline{P}+\overline{\rho} )+\frac{1}{4} \C\C_X f h r^2 X(3\overline{P}+\overline{\rho} )-p_3'-p_{11}'+p_1''+p_6+p_8\nonumber\\
&+p_{13}+p_{14},\;q_5=\C\C_X X fh \overline{P} r^2  +2 p_{13}+p_{14},\; q_6=2 p_{25}\frac{f r^2 X \C_X^2 }{h}\left(\frac{\overline{P}+\overline{\rho} }{c_m^2}-9 \overline{P}-\rho \right)\nonumber\\
& q_7=p_5+p_{12}+f r^2 X\varphi' \C_X^2 \left(\frac{1}{2} (9\overline{P}+\overline{\rho} )-\frac{\overline{P}+\overline{\rho} }{2 c_m^2}\right)+f r^2 \varphi'  \C_X \left(\frac{1}{2} \C (\overline{P}+\overline{\rho} )-\frac{\C (\overline{P}+\overline{\rho} )}{2 c_m^2}\right)
\nonumber\\
&q_8=2 p_{22}-\C\C_X\frac{2  f \overline{P} r^2  \varphi '}{h}.
\end{eqnarray}
\subsection*{$Y_i$:}
\begin{eqnarray}
&Y_0=\frac{f'}{f}-\frac{h'}{h}+\frac{2}{r},\quad Y_1 =\frac{e_8+2 q_3-(e_2/e_4)\left(e_{10}+2 q_5\right)}{2 q_0 }\nonumber\\
&Y_2=\frac{\left(\left(e_6+2 q_1\right)-(e_0/e_4) \left(e_{10}+2 q_5\right)\right) f'}{ f q_0}+\frac{e_9+2 q_4-(e_3/e_4) \left(e_{10}+2 q_5\right)}{2 q_0},\nonumber\\
& Y_3=\frac{4 \FK_X \varphi'}{\cR }\frac{  e_6+2 q_1-(e_0/e_4) \left(e_{10}+2 q_5\right)}{2 q_0}.
\end{eqnarray}
\subsection*{$Z_i$:}
\begin{eqnarray}
&Z_0=\frac{\left(q_7+q_8\right)-\left(e_0+e_1\right) (q_8'/e_4)}{ q_6},\quad Z_1 =\frac{ q_6'-q_8'(e_2 /e_4)}{q_6}\nonumber\\
&Z_2=\frac{2f' \left(  q_8-  q_8'(e_0/e_4)\right)}{f q_6}+\frac{q_7'-q_8'(e_3/e_4)}{q_6},\quad Z_3=\frac{4 \FK_X \varphi'}{\cR }\frac{ \left(q_8- q_8'(e_0/e_4)\right)}{q_6}-\frac{2 p_{23}}{q_6}.
\end{eqnarray}
Note that $f'$, $h'$ and $\overline{P}'$  can be eliminated by using the background equations (\ref{eh}), (\ref{ef}) and (\ref{eP}) and the scalar field is replaced by substituting (\ref{edphi}).

\bibliographystyle{ieeetr}
\bibliography{bibliography}

\end{document}